# Pyramidal Columns Acting as Attention Heads

Fabrizia Giulia Garavaglia[1,2] and Marco Giunti[3]

## Abstract

Can the functional architecture of a Transformer attention head be given a biological construction? More specifically, could a suitable group of pyramidal columns in posterior temporal language regions (PTLRs) constitute the biological construction of a neuromorphic mechanical realization of an attention head, under a mathematical instantiation that need not coincide with its standard Transformer instantiation? This paper investigates this hypothesis by examining possible correspondences between the functional organization of attention heads and neurophysiological mechanisms operating within and across pyramidal columns involved in human language processing. Our analysis builds on a mechanical-structural description of Transformer attention heads as systems of simple neural networks linked through output-input and output-weight interconnections. We examine the functional organization of an attention head and the input-dependent transformations it produces, and ask how the corresponding processing relations could be constructed in biological neural tissue. We then identify candidate neural mechanisms of signal transmission and modulation within and across pyramidal columns that may jointly support such a construction. On this basis, we propose that a suitably organized group of pyramidal columns in PTLRs could constitute a biological construction of a neuromorphic mechanical realization of an attention head. Whether the required functional organization actually exists in PTLRs remains an open empirical question. The hypothesis does not require the biological system to reproduce the standard mathematical instantiation of Transformer attention: the same functional architecture may be mechanically realized and mathematically instantiated in a different way, consistent with its biological construction.

## 1. Introduction

A central question raised by the mechanical-structural analysis of Transformer processing is whether the functional organization identified in artificial systems can also receive alternative physical constructions. This question becomes particularly interesting in the case of attention heads. In previous work, Transformer attention heads were redescribed as systems of simple neural networks organized through output-input and output-weight interconnections, the latter allowing outputs generated during inference to determine the weights, and hence the transformations, of other networks (Giunti & Garavaglia, 2026a). The present paper asks whether an attention head, understood at this level of functional organization, could also receive a biological construction.

---

[1] Dipartimento di Pedagogia, Psicologia, Filosofia, Università degli Studi di Cagliari.
[2] Corresponding author, fabriziag.garavaglia@unica.it.
[3] ALOPHIS – Applied LOgic, Philosophy, and HIstory of Science, Università degli Studi di Cagliari.

More specifically, we investigate the hypothesis that a suitably organized group of pyramidal columns in posterior temporal language regions (PTLRs), traditionally associated with Wernicke's area, could constitute the biological construction of a neuromorphic mechanical realization of an attention head. Here, an attention head is considered as an abstract system type, that is, in terms of its functional architecture independently of any particular mathematical instantiation (Giunti & Garavaglia, 2026b, §§ 7.1–7.2, 7.5, 8.4). A mechanical realization specifies an idealized system of components and interactions that preserves this functional organization; a neuromorphic realization does so in terms of neural-like processing components and interconnections. Treating the attention head itself as an abstract system type allows the question of its possible biological construction to be posed at this modular level, without requiring a corresponding construction of the Transformer as a whole. The mathematical instantiation assigned to such a realization need not coincide with the standard mathematical instantiation of Transformer attention: what matters is that it preserve the relevant functional architecture while being consistent with the biological construction under consideration.

This question is motivated by a specific feature of Transformer inference. During training, the model acquires matrices and vectors of parameters whose values remain fixed during a given inference episode. During inference, however, the processing of the current input also generates parameter structures whose values depend on that input and which in turn define further, dynamically constructed transformations of token representations. In the mechanical-structural description adopted here, this organization is made explicit through output-weight interconnections, whereby the output of one processing component determines how another component transforms its input. This suggests a possible point of contact with biological neural processing: input-dependent transformations might be constructed through local mechanisms of signal modulation, in which neural activity modifies the way other signals are transmitted or integrated. The biological problem we address is therefore not whether cortical tissue reproduces the standard matrix operations of Transformer attention, but whether neural mechanisms can construct the same relevant functional organization through different physical mechanisms and under a different mathematical instantiation.

Recent research has highlighted the potential of neuromorphic principles for developing alternative realizations of Transformer-like processing (Leroux, Finkbeiner, & Neftci, 2025). An example is the analog in-memory computing architecture for self-attention proposed by Leroux, Manea, et al. (2025), aimed at improving computational efficiency. Other studies have investigated correspondences between Transformer attention and behavioral and neural measures of human language processing (Gao et al., 2025). The present study addresses a different but related question: whether biological neural circuitry could provide a construction of a neuromorphic mechanical realization of the functional architecture of an attention head, potentially under a mathematical instantiation different from the standard Transformer one.

This perspective directs attention to neural mechanisms of signal transmission and modulation. Pyramidal neurons and interneurons within cortical columnar circuitry provide mechanisms of signal transmission and modulation that may support dynamically configurable processing relations. We therefore focus on pyramidal-column circuitry in PTLRs as the main candidate for such a biological construction. We do not assume that pyramidal columns already exhibit the functional organization of attention heads. Rather, we investigate whether suitably organized groups of such cortical microcircuits could collectively constitute a biological construction of that organization.

The distinction between relatively persistent neural organization and activity-dependent modulation also suggests a possible comparison with the distinction between training and inference in artificial systems. Training determines the relatively stable parameter structure subsequently used during Transformer inference, whereas biological neural organization reflects evolutionary history and is further shaped throughout an individual's life by developmental processes, including maturation and experience-dependent plasticity associated with learning. During ongoing language processing, by contrast, neural activity can transiently modify how signals are transmitted and integrated. We use this comparison only as a heuristic organizational distinction: no one-to-one identification between Transformer training and any single biological developmental process is assumed.

The broader motivation is twofold. First, determining whether biological neural tissue can construct the relevant functional organization may contribute to understanding the range of physical constructions available for a neuromorphic mechanical realization of attention. Second, a biological construction represents only one possible physical construction of such a realization. The same neuromorphic mechanical realization may, under a given mathematical instantiation, admit different physical constructions; it may also be assigned different mathematical instantiations, each of which may in turn guide different physical constructions. Investigating the biological case may therefore help clarify which aspects of the proposed organization belong to the neuromorphic mechanical realization itself and which depend on a particular mathematical instantiation or physical construction. These broader implications remain secondary, however, to the specific aim of the present paper: evaluating the hypothesis that groups of pyramidal columns in PTLRs could provide a biological construction of a neuromorphic mechanical realization of an attention head.

The paper proceeds as follows. Section 2 introduces the relevant organization of posterior temporal and frontal language regions and situates the proposed biological construction within this broader cortical system. Section 3 presents the functional organization of the attention head and introduces its neuromorphic mechanical realization, together with pyramidal-column circuitry as the main candidate for its biological construction. Section 4 examines neural mechanisms of signal transmission and modulation that could construct the functional relations made explicit by output-weight interconnections. Section 5 develops the proposed biological construction in greater detail by relating the components and interconnections of the neuromorphic realization to candidate neural mechanisms. Section 6 assesses the scope and limitations of the proposal and its broader implications for alternative physical constructions of attention.

## 2. A hypothetical mapping between Transformer modules and human language-processing areas

To situate the present study within a broader framework, we begin by sketching a hypothetical mapping between the functional modules of the Transformer and cortical regions involved in language processing. We restrict the discussion to those regions most directly involved in language-related functions, since the standard Transformer considered here is itself a system for linguistic processing.

The classical Wernicke–Geschwind model described language processing in terms of relatively circumscribed posterior and frontal language areas, traditionally identified with Wernicke's and Broca's areas, respectively (Geschwind, 1970; Dronkers et al., 2007). From a more distributed perspective, these areas can be understood as parts of broader posterior temporal language regions (PTLRs) and frontal language regions (FLRs) (Tremblay & Dick, 2016). Both PTLRs and FLRs receive and integrate information originating in other cortical and subcortical regions through auditory and other sensory-related pathways. Some of this information may reach posterior temporal and frontal regions through partially parallel routes, so that sensory input to FLRs need not first be processed by PTLRs (Rocchi et al., 2021). Communication between temporal and frontal language regions is likewise supported by multiple dorsal and ventral pathways and is therefore not mediated exclusively by the arcuate fasciculus (Friederici, 2015). Nevertheless, parts of the PTLRs and FLRs are reciprocally connected through the arcuate fasciculus (Giampiccolo & Duffau, 2022). FLRs also project to premotor and motor cortical areas involved in speech production and motor control. More generally, both PTLRs and FLRs participate in a wider network of afferent and efferent connections with other cortical and subcortical regions.

In the usual functional description of a decoder-only Transformer, a token sequence is first converted into initial vector representations through embedding and positional encoding and then passed to the multi-head self-attention module of the first Transformer layer (Vaswani et al., 2017). Within this module, the attention heads process the same input representations in parallel, each producing a new representation for every token. For each token, the outputs of the different attention heads are concatenated and then processed by the remaining operations of the layer. The resulting token representations constitute the input to the next Transformer layer, where the same sequence of processing steps is repeated. This pattern continues through the final layer, whose output is passed to the forecast module, which maps the resulting token representations onto probability distributions over the model vocabulary.

The same functional organization of the Transformer stack can, however, be rearranged without representing the $N$ layers as $N$ successive blocks. In the mechanical-structural description adopted here, the operations of each layer can instead be distributed between two main areas connected in a serpentine pattern. Area 1 contains only the $h$ attention heads of each of the $N$ layers. Area 2 contains all the remaining operations of each layer, beginning with the dense linear networks corresponding to the output projection of the multi-head attention module and continuing through the first residual addition and mean-and-variance normalization and optimization,[4] the feed-forward network module, and then the second residual addition and mean-and-variance normalization and optimization, up to the one-to-one additive networks corresponding to its bias parameters. The attention heads of the first layer receive their input from the embedding and positional-encoding module, which is located outside these two areas. For every subsequent layer, the input is the final output of the preceding layer: one branch enters Area 1 and feeds the $h$ attention heads, while the other remains in Area 2 and provides the residual input to the first residual addition. Accordingly, for every layer except the last, the final output produced in Area 2 bifurcates in this way. The output of the final layer neither bifurcates nor returns to Area 1; it is sent directly to the subsequent forecast module, which is likewise located outside these two areas.

With this rearrangement, the two areas are connected bidirectionally. At the boundary between them, the outputs of the $h$ attention heads of each layer provide $n_{\max}\, h\, d_v$ channels from Area 1 to Area 2, for a total of $N\, n_{\max}\, h\, d_v$ channels for all the $N$ layers. In the opposite direction, the $n_{\max}\, d_{\text{model}}$ channels carrying the output of each layer return from Area 2 to Area 1 only between successive layers; since the final layer instead sends its output to the forecast module, this yields $(N - 1)n_{\max}\, d_{\text{model}}$ channels from Area 2 to Area 1. The two directions contain the same number of cross-area interconnections exactly when

$$N\, n_{\max}\, h\, d_v \;=\; (N-1)n_{\max}\, d_{\text{model}}$$

that is, after cancelling $n_{\max}$, when

$$h\, d_v \;=\; (1 - \frac{1}{N})\, d_{\text{model}}$$

---

[4] Here, following TR1, *mean-and-variance normalization and optimization* denotes the complete module comprising normalization followed by an affine transformation with learned scaling and bias parameters. The latter adjust the variance and mean, respectively, of the normalized representations without any parameter updating during inference.

Hence, whenever

$$h\, d_v \;>\; (1 - \frac{1}{N})\, d_{\text{model}}$$

the number of interconnections from Area 1 to Area 2 is greater than the number in the reverse direction. Since standard Transformer architectures typically satisfy $h\, d_v \;=\; d_{\text{model}}$, this inequality is automatically satisfied for every finite $N \;>\; 1$, and even more generally whenever $h\, d_v \;\geq\; d_{\text{model}}$. The asymmetry predicted by the model suggests a testable neuroanatomical hypothesis: axonal projections through the arcuate fasciculus may be more numerous from PTLRs to FLRs than in the reverse direction.

On the basis of this broader correspondence, we now narrow our focus to Area 1 and to its possible biological counterpart. In the rearrangement described above, Area 1 contains the attention heads of all Transformer layers. We therefore investigate whether suitable neural circuitry within PTLRs—particularly circuitry organized around pyramidal columns—could provide a biological construction of a neuromorphic mechanical realization of these attention heads. Since parts of the PTLRs are reciprocally connected with FLRs through the arcuate fasciculus, this hypothesis also fits the broader two-area organization outlined above. Our analysis will consequently focus on the structure and signaling mechanisms of pyramidal-column circuitry in PTLRs and on their possible correspondence with the functional organization of attention heads.

For the purposes of this study, we use the term *pyramidal column* to designate a locally organized cortical microcircuit centered on pyramidal neurons and their interactions with interneurons, potentially involving connections across cortical depths as well as lateral connections with other microcircuits. This characterization does not presuppose sharply bounded anatomical columns or a single circuit extending through every cortical layer. The anatomical and functional status of cortical columns remains debated (Rockland, 2010; da Costa & Martin, 2010). Evidence of cross-laminar speech-related organization in human superior temporal cortex (Leonard et al., 2024) and of specialized synaptic interactions among pyramidal neurons in human temporal cortex (Planert et al., 2026) motivates the investigation of such circuitry, without establishing the specific organization hypothesized here.

- Pyramidal neurons are excitatory neurons and primarily use glutamate as their neurotransmitter.
- Interneurons are predominantly inhibitory and primarily use GABA as their neurotransmitter.
- Other neuronal types (e.g., spiny stellate cells) also contribute to local cortical processing.

We focus on pyramidal neurons and GABAergic interneurons because the two candidate mechanisms examined below involve GABAergic modulation of glutamatergic transmission and glutamate-dependent changes in dendritic-spine formation and synaptic efficacy. This focus does not exclude the involvement of other neuronal populations in the proposed biological construction. Pyramidal neurons are particularly relevant in this respect because their basal and apical dendritic compartments allow them to receive and integrate synaptic inputs from different sources (Spruston, 2008). In human temporal cortex, distinct pyramidal-neuron subtypes also exhibit differences in dendritic morphology and local synaptic connectivity (Planert et al., 2026).

A further feature that will be important for our analysis concerns the possible roles of glutamatergic signaling. In addition to carrying excitatory signals between neurons, glutamatergic activity can participate in processes that modify the efficacy or configuration of synaptic transmission. This distinction between signal transmission and signal modulation will become central when we examine possible biological counterparts of the different types of interconnection operating within an attention head. We return to these mechanisms in the following sections.

## 3. A neuromorphic mechanical realization of an attention head

The hypothetical mapping outlined in the previous section becomes informative only if the functional organization of an attention head is specified at a level that can be related to neural organization without requiring the biological system to reproduce the standard mathematical instantiation of Transformer attention. We therefore consider the attention head as an abstract system type and ask how its functional architecture can be given a neuromorphic mechanical realization. Such a realization specifies neural-like components and interconnections that preserve the relevant organization of transformations and functional roles, while leaving open the particular mathematical instantiation assigned to them (Giunti & Garavaglia, 2026b, §§ 7.3, 7.5, 8.4). The subsequent question is whether this neuromorphic organization can in turn guide the identification of a possible biological construction.

In the mechanical-structural description developed in previous work, an attention head can be represented as a system of simple neural networks interconnected through output-input and output-weight interconnections (Giunti & Garavaglia, 2026a, §3; Supplementary Material, Fig. S11c). In the standard mathematical instantiation, matrix-defined linear transformations are represented by dense linear networks, componentwise rescaling by one-to-one multiplicative networks, and Softmax by a distinct operational block. The initial query, key, and value transformations are defined by static weights, whereas later transformations within the head can receive dynamically generated weights. In particular, outputs produced along the key and value processing paths determine the weights of networks operating later in the computation. Output-weight interconnections therefore make explicit a central feature of attention processing: the output of one component does not necessarily become an input to another component, but may instead determine how that component transforms its input.

A neuromorphic mechanical realization makes the internal organization of these networks more explicit. Signal channels are organized as axon-like and dendrite-like channels around soma-like and synapse-like functional units. In dense networks, an incoming signal can be distributed through branching points of axon-like channels and reach multiplicative-role synapse-like units, while downstream soma-like units integrate the resulting signals. Static multiplicative-role synapse-like units have a single input channel and a fixed weight, whereas dynamic multiplicative-role synapse-like units have two input channels: one carries the signal to be transmitted and the other supplies a dynamically generated weight through an output-weight interconnection. One-to-one rescaling networks follow the same general organization without the branching required by dense networks.

Softmax-role operational blocks are represented by distinct functional units receiving their signals through synapse-like identity units. Here, *multiplicative* and *Softmax* designate functional roles rather than particular mathematical operations or functions. Figure 1 shows the resulting bidirectional realization without input. The complete bidirectional series and the detailed graphical conventions are provided in the *Supplementary Materials (SM)*, Legend and Figs. S1–S1e.

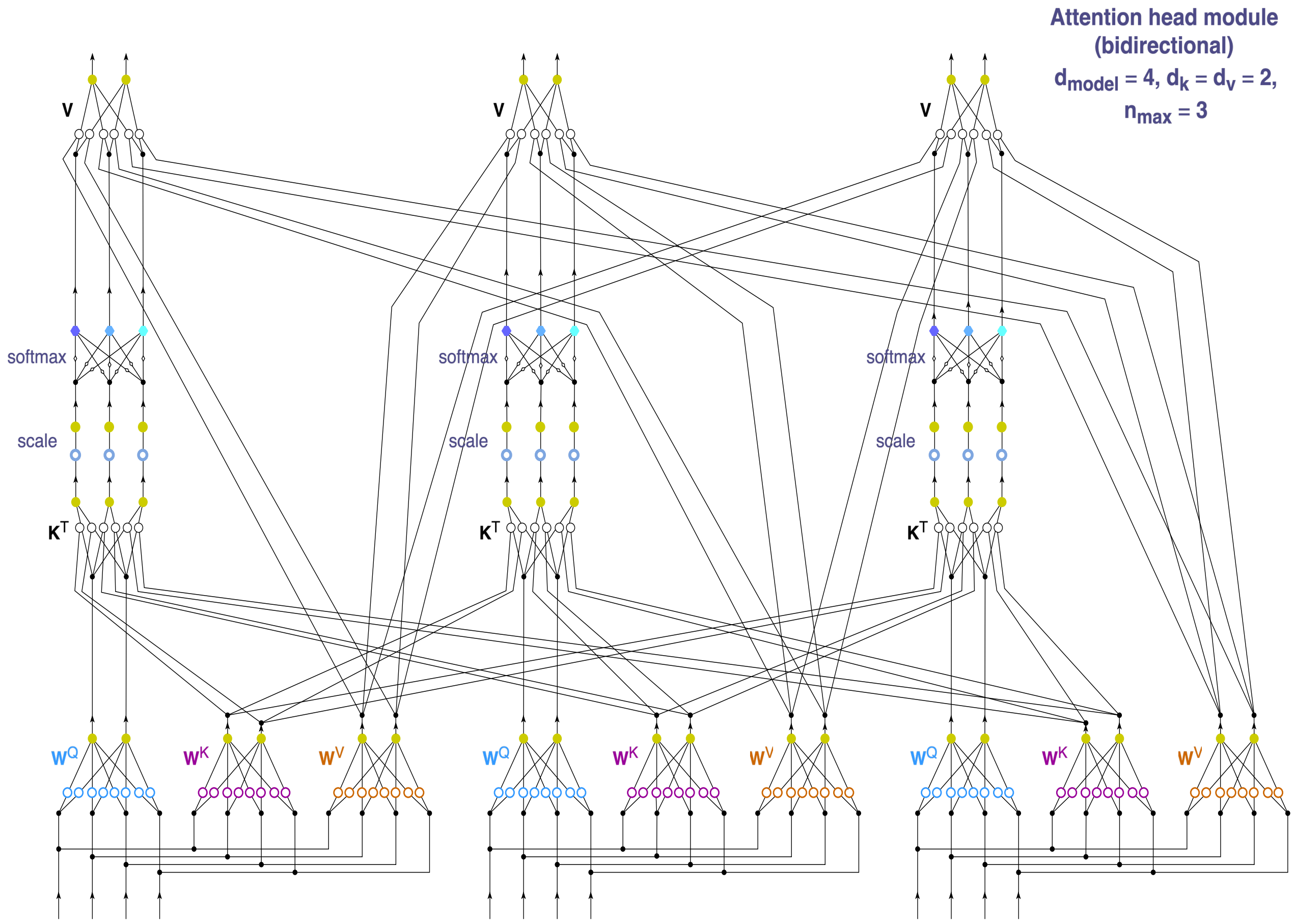

Figure 1. Neuromorphic mechanical realization of a bidirectional attention head with $n_{max} = 3, d_{model} = 4$, and $d_k = d_v = 2$, shown without input. The three token-processing streams are arranged in successive columns from left to right. For the graphical conventions and the corresponding activation figures, see SM, Figs. S1–S1e.

For decoder-only attention, we adopt an autoregressive realization in which causality is represented structurally rather than through an additive masking operation. The relevant restriction is built into the structural organization of the neuromorphic realization: token-indexed components have access only to the signals that are causally available to them, and the corresponding networks and interconnections are reduced accordingly. The mechanical realization therefore preserves the autoregressive functional organization of the attention head without requiring a separate masking network. This is important for the present purpose, because the biological construction need only reproduce the relevant causal organization, not the particular masking operation used in the standard mathematical instantiation of the Transformer. Figure 2 shows this autoregressive neuromorphic realization without input; its complete activation series is shown in SM, Figs. S2–S2e.

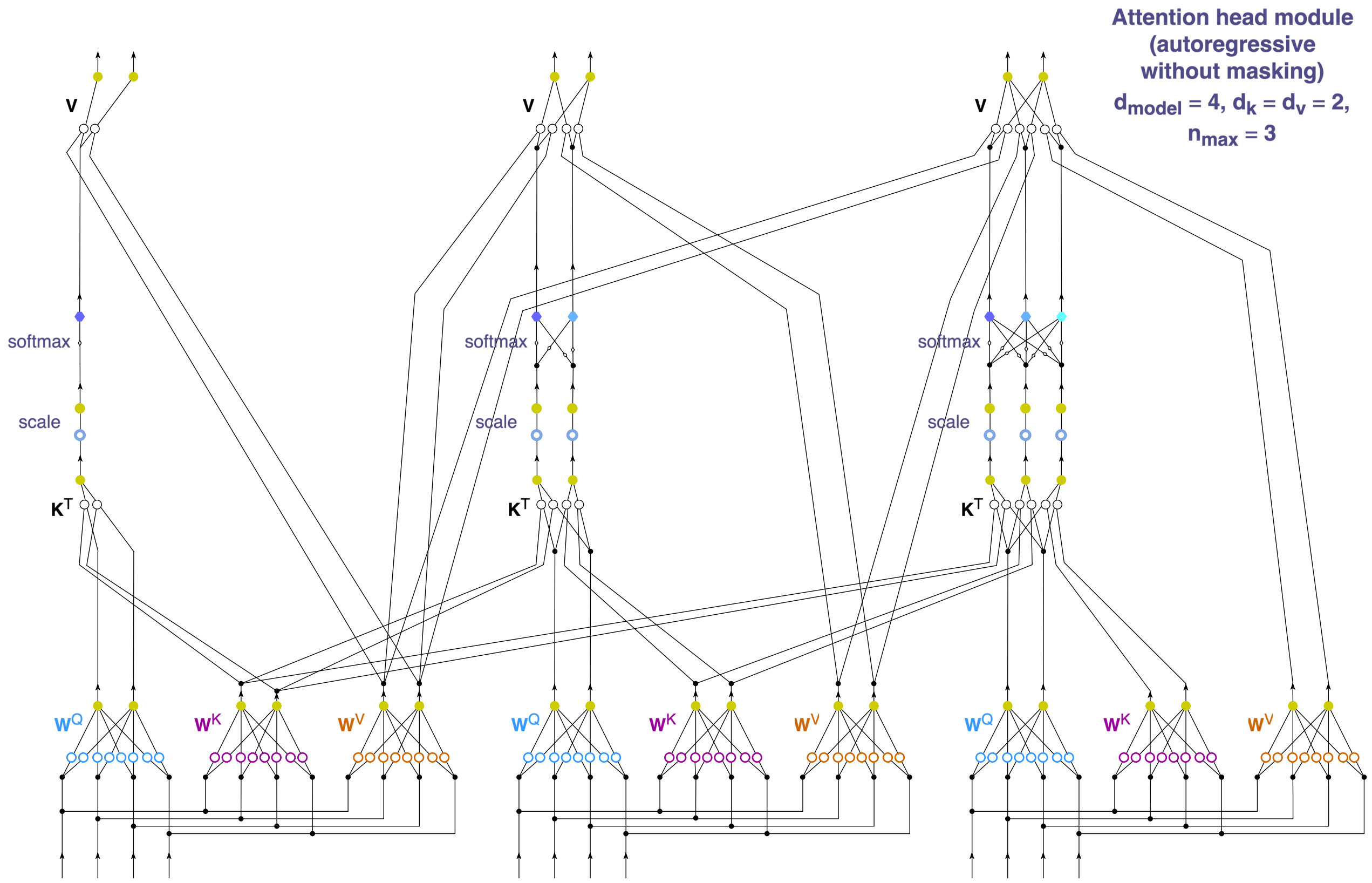

Figure 2. Neuromorphic mechanical realization of an autoregressive attention head without masking, with $n_{max} = 3, d_{model} = 4$, and $d_k = d_v = 2$, shown without input. Autoregressivity is implemented through the structural organization of the networks and their interconnections, rather than by a separate masking operation. For the graphical conventions and the corresponding activation figures, see SM, Figs. S2–S2e.

Figure 3 highlights a fundamental feature of the neuromorphic realization: the generation of dynamic weights and the application of the corresponding transformations are distinct but coordinated processes. In particular, the key- and value-processing paths can dynamically configure downstream networks independently of whether these networks actually process any input. This makes especially clear the functional distinction between output-weight and output-input interconnections. For completeness of the supplementary series, Fig. 3 is also reproduced in SM as Fig. S2e.

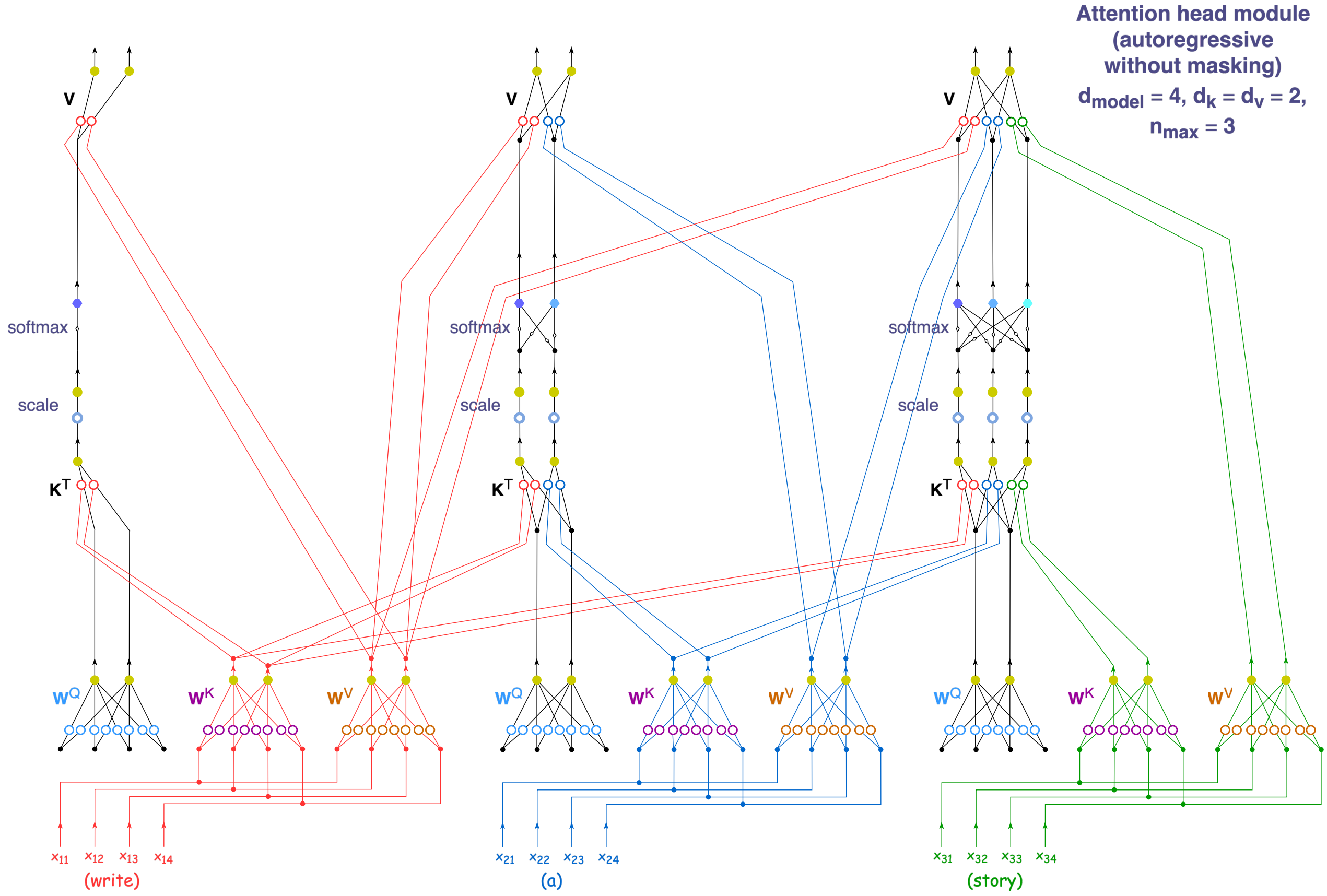

Figure 3. The autoregressive attention head in Fig. 2 activated by a three-token input, which is supplied only to the networks generating keys and values. The sole function of these signals is to generate the dynamic weights that are transmitted through output-weight interconnections to downstream multiplicative-role synapse-like units. Since no input is supplied to the query-generating networks, no output is produced by the attention head. The figure therefore isolates the weight-generating function of the key and value processing paths. For the graphical conventions and the complete autoregressive series, see SM, Figs. S2–S2e.

The neuromorphic representation also suggests a set of correspondences between its components and biological neural structures. Table 1 summarizes these correspondences. In particular, soma-like functional units can be associated with neuronal somata, axon-like channels with axons, branching points with the origins of axonal arborizations, and dendrite-like channels with dendrites. Multiplicative-role synapse-like units with a single input can be associated with ordinary synaptic transmission or, more precisely, with a synapse whose efficacy is locally fixed. Multiplicative-role synapse-like units with two inputs admit two relevant biological correspondences already identified in TR1: axo-axonic synapses provide a strong morphological and functional correspondence, since the modulatory relation is structurally explicit; neighboring axo-dendritic synapses on the same dendritic segment also provide a morphological and functional correspondence, although in this case morphology alone does not determine which synapse is modulatory and which is modulated, a distinction that must be established functionally (Giunti & Garavaglia, 2026a, §7.1; Supplementary Material, Fig. S20).

Table 1. Functional and morphological correspondences between components of the neuromorphic realization and candidate biological counterparts.

| Neuromorphic component | Functional role | Candidate biological counterpart | Type of correspondence |
|---|---|---|---|
| Integrative soma-like functional unit (yellow filled circle) | Integrates incoming signals; may also receive a single input, as in rescaling units. | Neuronal soma | Morphological and functional |
| Softmax-role functional unit (colored hexagon) | Compares the strength of the signal on its corresponding input channel with the combined strength of the signals on all input channels, thereby producing a normalized attention value. | Potentially, a neuronal circuit corresponding to a more detailed neuromorphic decomposition of the Softmax-role block. | Morphological and functional correspondence to be established at the circuit level. |
| Axon-like channel | Carries signals away from soma-like units or from neurons external to the represented system. | Axon | Morphological and functional |
| Branching point of a signal channel (small filled circle) | Distributes a signal along multiple branches. | Branching point of an axonal arborization | Morphological and functional |
| Dendrite-like channel | Carries a signal from a synapse-like unit toward a soma-like functional unit. | Dendrite | Morphological and functional |
| Static multiplicative-role synapse-like unit (colored hollow circle; gray for rescaling) | Transforms an incoming signal according to a fixed weight. | Synapse with locally fixed efficacy | Morphological and functional |
| Dynamic multiplicative-role synapse-like unit (hollow circle) | Receives the signal to be transmitted through its first input and a dynamically generated weight through its second input. | Axo-axonic synapse onto a presynaptic terminal; or local interactions between neighboring axo-dendritic synapses on the same dendrite. | Morphological and functional; which synapse modulates the other in the neighboring axo-dendritic case must be established functionally. |
| Synapse-like identity unit (diamond) | Transmits the incoming signal unchanged to a Softmax-role functional unit. | Neural circuitry, potentially involving electrical synapses. | Functional correspondence to identity transmission remains to be established at the circuit level; no one-to-one morphological correspondence is assumed. |

The Softmax-role block may admit a more detailed neuromorphic realization in terms of interconnected, non-integrative soma-like functional units and synapse-like identity units. Its possible biological construction should therefore be investigated at the level of neural circuitry, rather than presupposing a one-to-one correspondence between the functional units represented in the figures and individual biological neurons. Electrical synapses may contribute to such a construction by providing direct coupling between neurons, but they do not generally transmit signals unchanged, since electrical transmission can involve attenuation and filtering (Curti & O'Brien, 2016). Two questions therefore remain open: how the required identity-transmission role could be biologically constructed, possibly through more complex neural circuits, and how the functional units of the Softmax-role block could themselves be constructed from interconnected neurons. A necessary next step is to develop a detailed neuromorphic mechanical realization of the Softmax-role block, specifying how its mathematical operations can be decomposed into elementary functional units and their interconnections. The subsequent challenge is to investigate whether these units and their functional composition can be biologically constructed using different neuronal populations and synaptic mechanisms.

Output-weight interconnections are especially important for the proposed construction. At the functional level, they coordinate two different forms of signaling: one signal is transmitted onward and contributes to the activity of a downstream functional unit, whereas another does not function as an output-input signal but instead determines or modulates the efficacy of the relation through which the first signal is transmitted or transformed. The correspondences summarized above show that biological neural circuitry provides mechanisms capable, in principle, of constructing this organization. The more specific question addressed in the following sections is whether such mechanisms can be organized within pyramidal-column circuitry in PTLRs—or, where necessary, in reciprocally connected neural structures outside PTLRs—so as to provide a biological construction of the complete attention-head architecture.

## 4. Candidate biological mechanisms for output-weight interconnections

The correspondences identified in the previous section suggest two different ways in which biological neural circuitry could construct the modulatory relation characteristic of output-weight interconnections. In the first, a modulatory signal changes the efficacy of transmission through an already established synaptic pathway. In the second, a signal contributes to the formation and functional activation of a new synaptic contact. We investigate these possibilities through GABAergic axo-axonic modulation and glutamate-dependent dendritic-spine mechanisms, respectively, considering how they might operate within pyramidal-column circuitry in PTLRs.

Synaptic organization is not fixed once and for all. Both the efficacy of established synapses and patterns of synaptic connectivity can be modified by experience and learning (LeDoux, 2002). Computational models of cortical columns also show how experience-dependent synaptic plasticity can modify circuit dynamics (Moreni et al., 2025). Axonal varicosities can develop into presynaptic boutons, while dendritic spines can emerge and undergo structural changes in response to neural activity (Yuste, 2023). The establishment of a functionally effective synapse involves appropriate presynaptic and postsynaptic structures, neurotransmitter receptors, and molecular mechanisms of synaptic adhesion and stabilization. These considerations suggest a distinction between modulating transmission through an existing synapse and making a potential synaptic contact functionally effective, which will guide the analysis that follows.

### 4.1. GABAergic axo-axonic modulation

The first candidate mechanism involves GABAergic modulation of synaptic transmission between pyramidal neurons through axo-axonic contacts. Consider two glutamatergic pyramidal neurons, A and J, belonging to the first and second functional processing stages, respectively. Neuron A transmits an excitatory signal to neuron J through an established axo-dendritic synapse. A GABAergic interneuron R, in turn, forms an axo-axonic contact with the presynaptic axon terminal of neuron A, thereby modulating its transmission to J. In this arrangement, the interneuron acts on the efficacy of an existing synaptic connection, rather than transmitting an additional signal directly to the postsynaptic neuron. Figure 4 illustrates this mechanism.

Axo-axonic contacts provide a well-documented mechanism of neural modulation, and presynaptic GABAergic activity can reduce glutamate release by acting on receptors located at excitatory nerve terminals (Cover & Mathur, 2021; Ladera et al., 2008). The relevant modulation can therefore affect synaptic transmission without requiring the formation of a new synaptic contact. Whether the particular arrangement represented in Fig. 4 operates within pyramidal-column circuitry in PTLRs remains, however, an empirical question.

At the functional level, this mechanism provides a candidate biological construction of a dynamic multiplicative-role synapse-like unit. The glutamatergic signal from neuron A occupies the role of the signal to be transmitted, whereas the GABAergic signal from interneuron R occupies the role of the dynamically generated parameter governing its transmission efficacy. The corresponding axo-axonic modulation thus exhibits the organization characteristic of an output-weight interconnection: the output of one neuronal component determines how another component transmits its signal. This correspondence does not require synaptic modulation to implement ordinary multiplication, since the multiplicative role admits different mathematical instantiations.

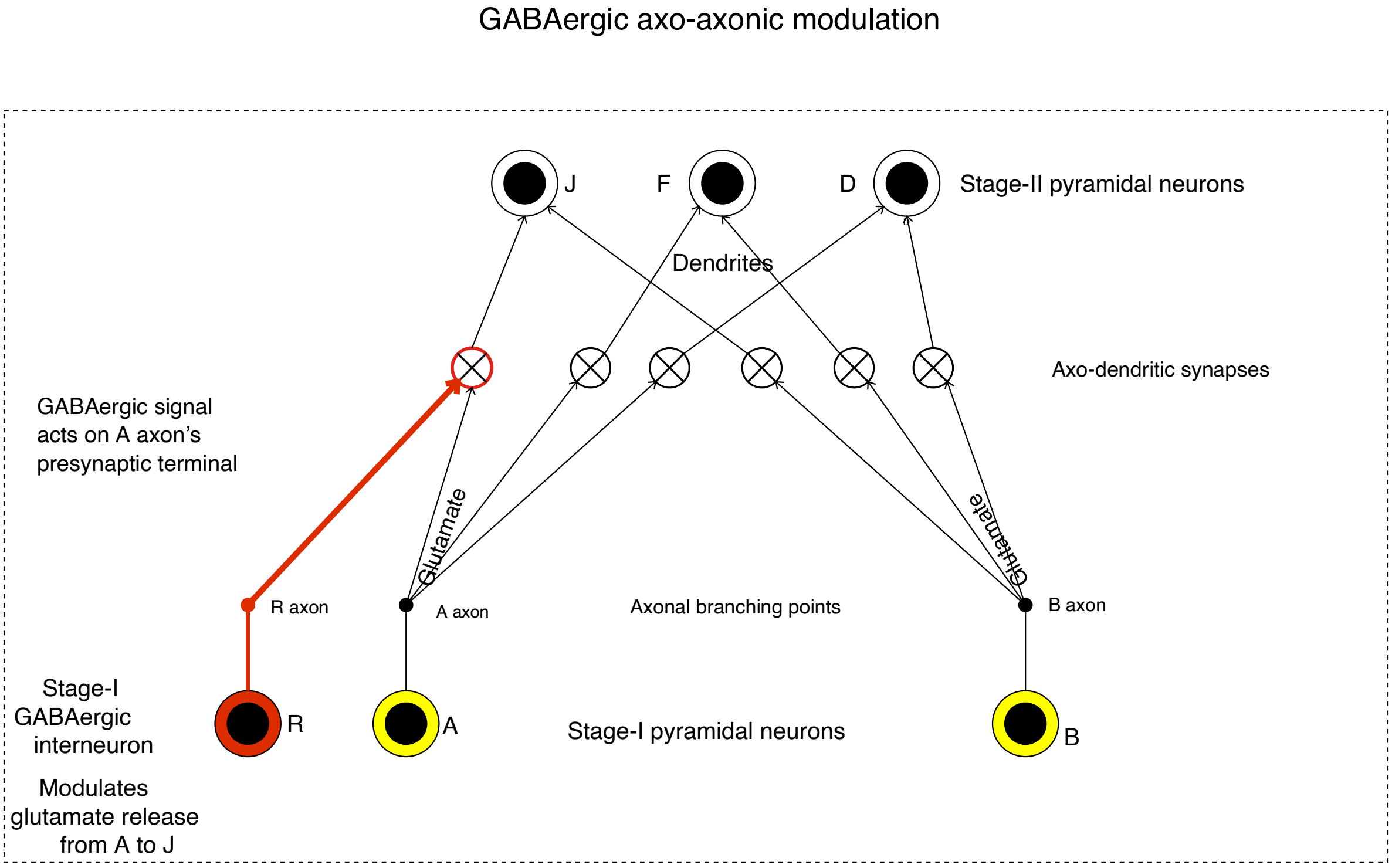

Figure 4. Candidate axo-axonic mechanism for the biological construction of an output-weight interconnection. A GABAergic interneuron (R) modulates glutamatergic transmission from a pyramidal neuron (A) to a postsynaptic pyramidal neuron (J) by acting on the presynaptic terminal of A.

## 4.2. Glutamatergic modulation and dendritic-spine formation

The second candidate mechanism involves local interactions between neighboring axo-dendritic synapses on the same dendrite, as already considered in previous work (Giunti & Garavaglia, 2026a, §7.1). In a simplified three-stage organization considered here, a pyramidal neuron of the first functional stage forms an established axo-dendritic synapse with a pyramidal neuron of the third stage. Glutamatergic activity at this synapse may promote the emergence of a nearby dendritic spine, providing a potential site for a second synaptic contact with the axon of a pyramidal neuron of the second stage. The first synapse thus plays a modulatory role, whereas the second, once functionally established, transmits the signal originating from the second-stage neuron. Although the established I–III synapse produces a local postsynaptic response, we hypothesize that this response is insufficient, by itself, to generate an output from the third-stage neuron. Its principal role in the proposed mechanism is to promote the formation and modulate the efficacy of the neighboring II–III synaptic contact.

Dendritic spines are dynamic structures whose formation and modification depend on neuronal activity (Yuste, 2023). Experimental studies have shown that local glutamate stimulation can induce rapid, spatially specific formation of new dendritic spines in developing cortical neurons. Such formation depends on NMDA receptor activation and intracellular signaling mechanisms, and the resulting spines can rapidly acquire functional properties (Kwon & Sabatini, 2011). Other studies have shown that newly formed spines can develop functional glutamatergic responses and establish structurally mature synaptic contacts within a relatively short time (Zito et al., 2009).

The emergence of a dendritic spine provides a potential site for synaptic transmission. The establishment of an effective synaptic contact also involves appropriate presynaptic and postsynaptic organization, including interactions between synaptic adhesion molecules such as presynaptic neurexins and postsynaptic neuroligins, together with functional glutamate receptors, notably AMPA and NMDA receptors (Dean & Dresbach, 2006). In the mechanism proposed here, a spine whose formation is promoted by glutamatergic activity at the established I–III synapse may form a contact with the axon of a second-stage neuron. This contact may become functionally effective when the required molecular and structural conditions are satisfied and the second-stage neuron supplies a glutamatergic signal.

At the functional level, this mechanism provides a second candidate biological construction of a dynamic multiplicative-role synapse-like unit. Glutamatergic activity at the established I–III synapse contributes to determining the efficacy of the synaptic relation through which the second-stage neuron transmits its signal to the third-stage neuron. The corresponding relation therefore has the functional organization of an output-weight interconnection. The precise dependence of synaptic efficacy on the two glutamatergic signals remains to be established, as does the possibility that the proposed sequence operates within PTLR circuitry on the temporal scale relevant to language processing. In §5, we examine how the two mechanisms described above could be coordinated within a pyramidal-column circuit hypothesized to constitute the biological construction of a neuromorphic mechanical realization of an attention head.

## 5. A candidate pyramidal-column construction of an attention head

We now examine how the two mechanisms discussed in §4 could be coordinated within pyramidal-column circuitry to provide a biological construction of a neuromorphic mechanical realization of an attention head. The proposed organization involves glutamatergic pyramidal neurons and GABAergic interneurons distributed across interconnected neural groups, with different populations contributing to signal transmission and modulation. Figure 5 illustrates a simplified circuit intended to make these relations explicit. The construction is investigated at the level of functional organization, without requiring biological neurons and synapses to reproduce the standard mathematical operations of Transformer attention.

For the purposes of this model, we distinguish three functional processing stages, labeled I, II, and III. These stages should not be identified with the anatomical layers of the cerebral cortex. In the first stage, three groups of neurons receive input-related activity and perform roles corresponding, respectively, to the query, key, and value processing paths of the attention head. The first group consists of glutamatergic pyramidal neurons that transmit signals toward the second stage. The second consists of GABAergic interneurons whose outputs modulate these transmissions, thereby occupying the role of dynamically generated parameters associated with the key-processing path. The third consists of glutamatergic pyramidal neurons whose outputs modulate synaptic transmission toward the third stage, in correspondence with the value-processing path. The assignment of these roles to different neuronal populations is a hypothesis concerning the biological construction of the functional architecture.

The first modulatory mechanism operates between stages I and II. Glutamatergic pyramidal neurons of stage I transmit signals to pyramidal neurons of stage II through established axo-dendritic synapses. GABAergic interneurons of stage I form axo-axonic contacts on the presynaptic terminals of these glutamatergic pathways, modulating the efficacy of their transmission as described in §4.1. The activity of the stage-II neurons consequently depends on both the incoming glutamatergic signals and the GABAergic modulation. In the proposed functional correspondence, these relations contribute to constructing the dynamically weighted processing associated with the interaction between queries and keys.

The second modulatory mechanism operates through local interactions between neighboring axo-dendritic synapses on the dendrites of stage-III pyramidal neurons. Glutamatergic pyramidal neurons belonging to the third group of stage I form established synapses with these neurons. Their activity produces local postsynaptic responses that, according to our hypothesis, promote the emergence of dendritic spines near the established contacts. The resulting spines provide potential sites for synaptic connections with axons originating from stage-II neurons. Although the established I–III synapses produce local postsynaptic responses, we assume that these responses alone are insufficient to generate an output from the stage-III neurons. Their principal role in the proposed organization is modulatory.

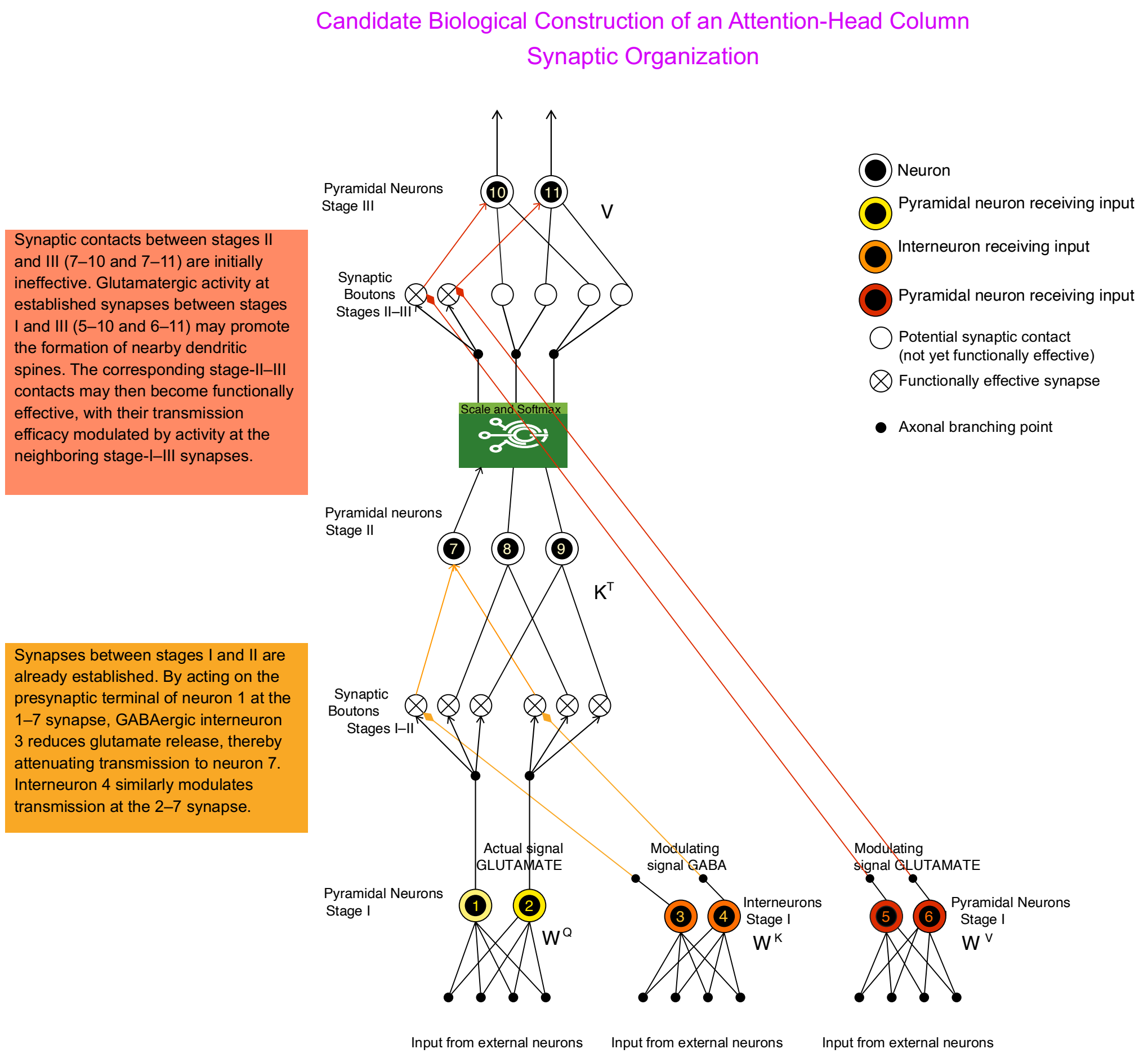

Figure 5. Candidate pyramidal-column circuit for the biological construction of a neuromorphic mechanical realization of an attention head. Three groups of first-stage neurons perform roles associated with query, key, and value processing. GABAergic interneurons modulate established synapses between stages I and II, while glutamatergic pyramidal neurons contribute to the configuration of synaptic transmission between stages II and III. The stages represent functional processing levels rather than anatomical cortical layers. Additional molecular and cellular mechanisms, including neurexin–neuroligin interactions and processes involving AMPA and NMDA receptors, contribute to the establishment of functionally effective synaptic contacts but are not represented in the figure.

Figure 6 represents an intermediate condition in which glutamatergic modulation has promoted the emergence of dendritic spines on stage-III neurons, while the corresponding II–III contacts have not yet become functionally effective. At the same time, GABAergic interneurons may already have modified the efficacy of established I–II synapses. The two mechanisms can therefore operate in a coordinated manner: one regulates the transmission of signals toward the second stage, while the other contributes to configuring the pathways through which the resulting signals can reach the third stage.

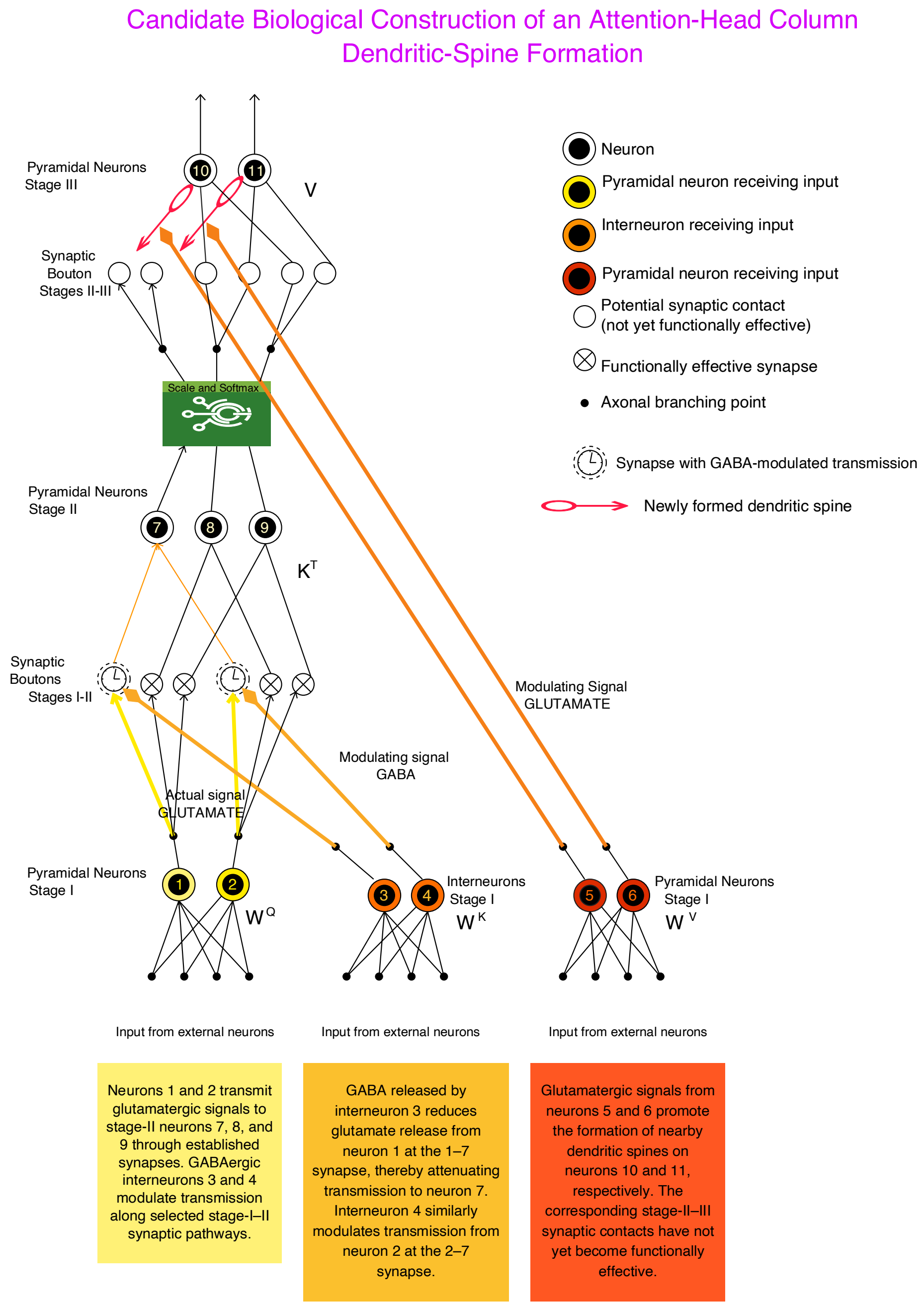

Figure 6. Intermediate configuration of the candidate circuit. GABAergic modulation affects established I–II synapses, while glutamatergic activity at established I–III synapses promotes the formation of dendritic spines near potential II–III contacts. The newly available contacts are not yet represented as functionally effective.

The subsequent condition is represented in Fig. 7. When the appropriate presynaptic and postsynaptic structures, adhesion mechanisms, and glutamate receptors are functionally organized, the dendritic spines may participate in effective synaptic contacts with stage-II axons. In the proposed mechanism, the efficacy of these contacts depends on the glutamatergic activity associated with both the established I–III synapses and the newly established II–III synapses. The signals reaching stage III therefore depend on the preceding processing at stage II and on the modulatory activity originating in stage I. The resulting activity of stage-III pyramidal neurons constitutes the candidate biological counterpart of the output generated by the corresponding processing paths of the attention head.

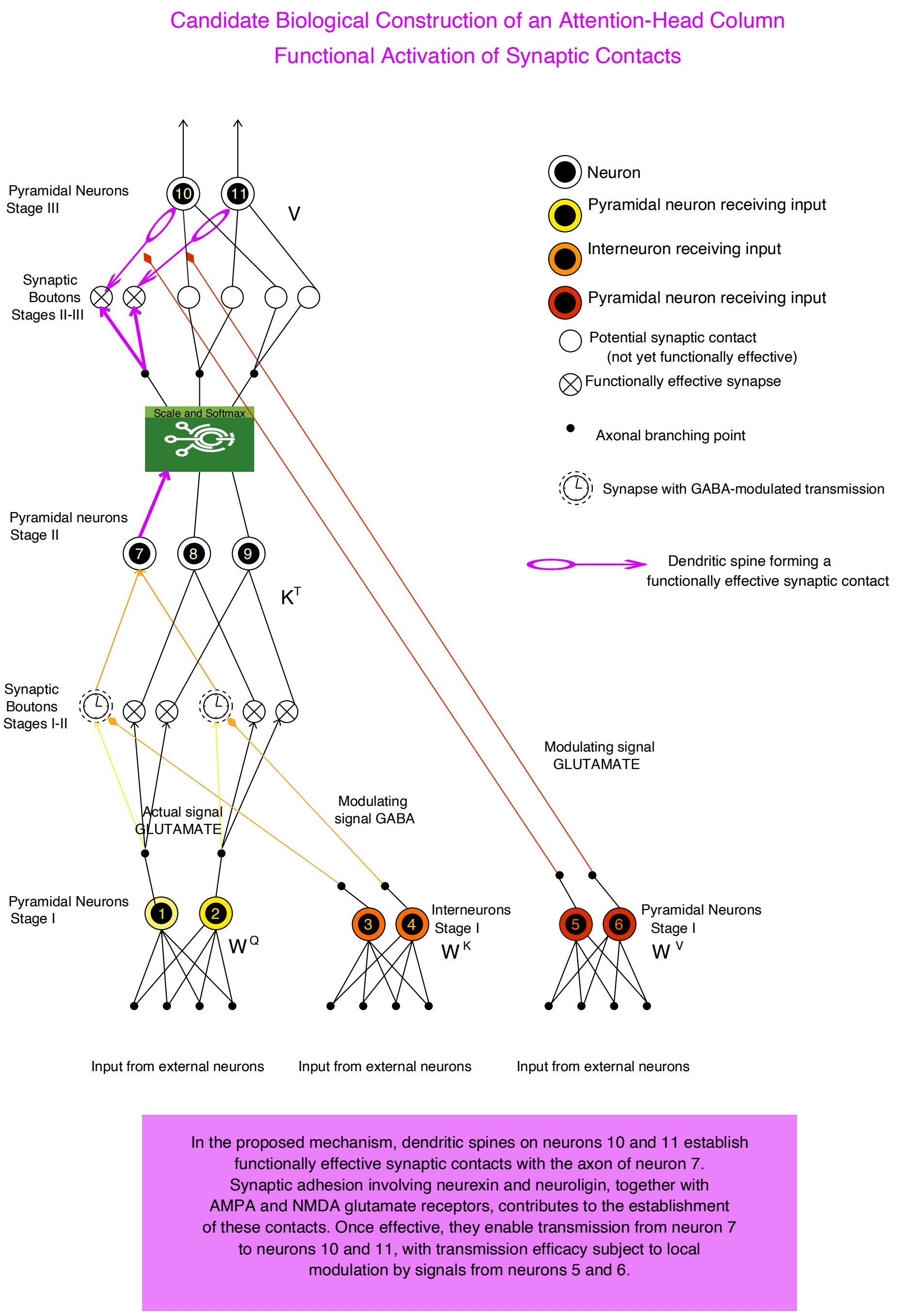

Figure 7. Subsequent configuration of the same circuit. The II–III contacts 7–10 and 7–11 have become functionally effective, allowing stage-II signals to contribute to the activity of stage-III pyramidal neurons. Their transmission efficacy is hypothesized to depend on local glutamatergic modulation originating from established I–III synapses.

The circuit represented in Figs. 5–7 is a simplified illustration of the processing relations involved in an attention head. In particular, the rescaling operations and the Softmax-role block that intervene between the two dynamically weighted transformations are not developed at the same biological level of detail. Their functional roles remain part of the neuromorphic mechanical realization described in §3. A complete biological construction would consequently require appropriate neural circuitry for these operations, including the comparison and normalization of signals across the available token positions. As discussed in §3, a more detailed neuromorphic decomposition of the Softmax-role block is possible in principle, but its biological construction has yet to be established.

The circuit represented in Figs. 5–7 corresponds to the first token-processing column of a bidirectional attention head with $n_{max} = 3$, $d_k = d_v = 2$, and $d_{model} = 4$. At the level of the simplified biological representation adopted here, the complete attention head can be constructed by placing two additional, identical columns alongside the first and introducing the appropriate axonal connections from stage-I modulatory neurons to the synaptic pathways of stages II and III, both within and across columns. These connections support the two biological modulation mechanisms described in §4 and provide candidate biological constructions of the output-weight interconnections represented in the neuromorphic mechanical realization of the complete bidirectional attention head shown in Fig. 1.

In the autoregressive case, by contrast, the token-processing columns are not identical: the number of neurons in stage II is one, two, and three in the first, second, and third columns, respectively. Moreover, the axonal connections responsible for modulation are causally restricted: they extend either within the same column or from earlier columns to subsequent ones, never in the reverse direction. This organization provides the biological counterpart of the structural restrictions characterizing the autoregressive neuromorphic mechanical realization shown in Fig. 2, without requiring an explicit masking operation.

Taken together, the GABAergic and glutamatergic mechanisms provide candidate biological constructions of the two types of dynamic synaptic modulation required by the proposed organization. Their coordination within pyramidal-column circuitry suggests a possible route toward the biological construction of an entire attention head, rather than merely of isolated multiplicative-role synapse-like units. Whether the necessary neural organization exists in PTLRs, whether the remaining functional components can be constructed, and whether the proposed glutamate-dependent synaptic sequence can operate on the temporal scale of language processing remain open empirical questions. The present model is therefore a specific, biologically motivated construction hypothesis, whose individual mechanisms and overall organization require further investigation.

## 6. Conclusion

This paper has presented a neuromorphic mechanical realization of the functional architecture of an attention head, making explicit the role of output-weight interconnections in both bidirectional and autoregressive processing. It has then identified two candidate biological mechanisms for constructing these interconnections—GABAergic axo-axonic modulation and local glutamatergic interactions between neighboring axo-dendritic synapses—and proposed how they could be coordinated within pyramidal-column circuitry in posterior temporal language regions (PTLRs). The resulting hypothesis concerns the biological construction of a neuromorphic mechanical realization of an attention head, potentially under a mathematical instantiation different from the standard Transformer one.

Further research should investigate whether the required neural organization exists in PTLRs and whether the proposed mechanisms, particularly the glutamate-dependent formation and functional activation of synaptic contacts, can operate on the temporal scale of language processing. It should also examine possible biological constructions of the rescaling operations and the Softmax-role block, as well as the coordination of modulatory signaling across pyramidal columns. Beyond these neurobiological questions, the proposed approach may provide a basis for exploring alternative physical constructions of dynamic attention, including neuromorphic hardware, and for investigating the possible relevance of such processing mechanisms to human language functions and their impairments.

# SUPPLEMENTARY MATERIALS

## LEGEND

Figures S1–S1e and S2–S2e show neuromorphic mechanical realizations of an attention head with $n_{max} = 3$, $d_{model} = 4$, and $d_k = d_v = 2$. Figures S1–S1e represent a bidirectional attention head, whereas Figs. S2–S2e represent an autoregressive attention head without masking. Token-processing streams are arranged in successive columns from left to right. Figures S1 and S2 show the corresponding structures without input; the lettered figures show their activation under different input conditions.

Yellow filled circles represent soma-like functional units; small filled circles represent branching points of signal channels. Hollow circles represent multiplicative-role synapse-like functional units. Small diamonds represent synapse-like identity units that transmit their input unchanged to the Softmax-role functional units, represented by small colored hexagons. Signal paths before synapse-like units or after soma-like units are axon-like channels, whereas those between synapse-like units and soma-like functional units are dendrite-like channels.

Multiplicative-role synapse-like units are of two kinds. Static synapse-like units have a single input channel and a fixed weight. Those whose weights are fixed through training are represented by colored hollow circles. The multiplicative-role synapse-like units implementing the rescaling operation are instead represented by thicker gray hollow circles, since their weight is fixed by the Transformer architecture rather than through training. Dynamic synapse-like units have two input channels: the first carries the signal to be transmitted, whereas the second supplies the weight. They are black when no weight-bearing signal is received through the second input and acquire the color of that signal when a dynamic weight is supplied through an output-weight interconnection.

In the activation figures, colored channels indicate signals currently propagating through the system; channels carrying no signal remain black. Signals passing through static synapse-like units retain their token-specific color. When a signal passes through a dynamic synapse-like unit whose weight has been supplied by a second input, the outgoing signal acquires the color of that dynamic weight. This convention makes the token-relative organization of attention visually explicit, in particular in the outputs of the $K^{\top}$ networks and in the subsequent dynamically weighted processing.

Here, *multiplicative* and *Softmax* designate functional roles rather than particular mathematical operations or functions. In the standard Transformer these roles are instantiated, respectively, by ordinary multiplication and the Softmax function; alternative mathematical instantiations may preserve the same functional architecture while assigning different operations or functions to these roles.

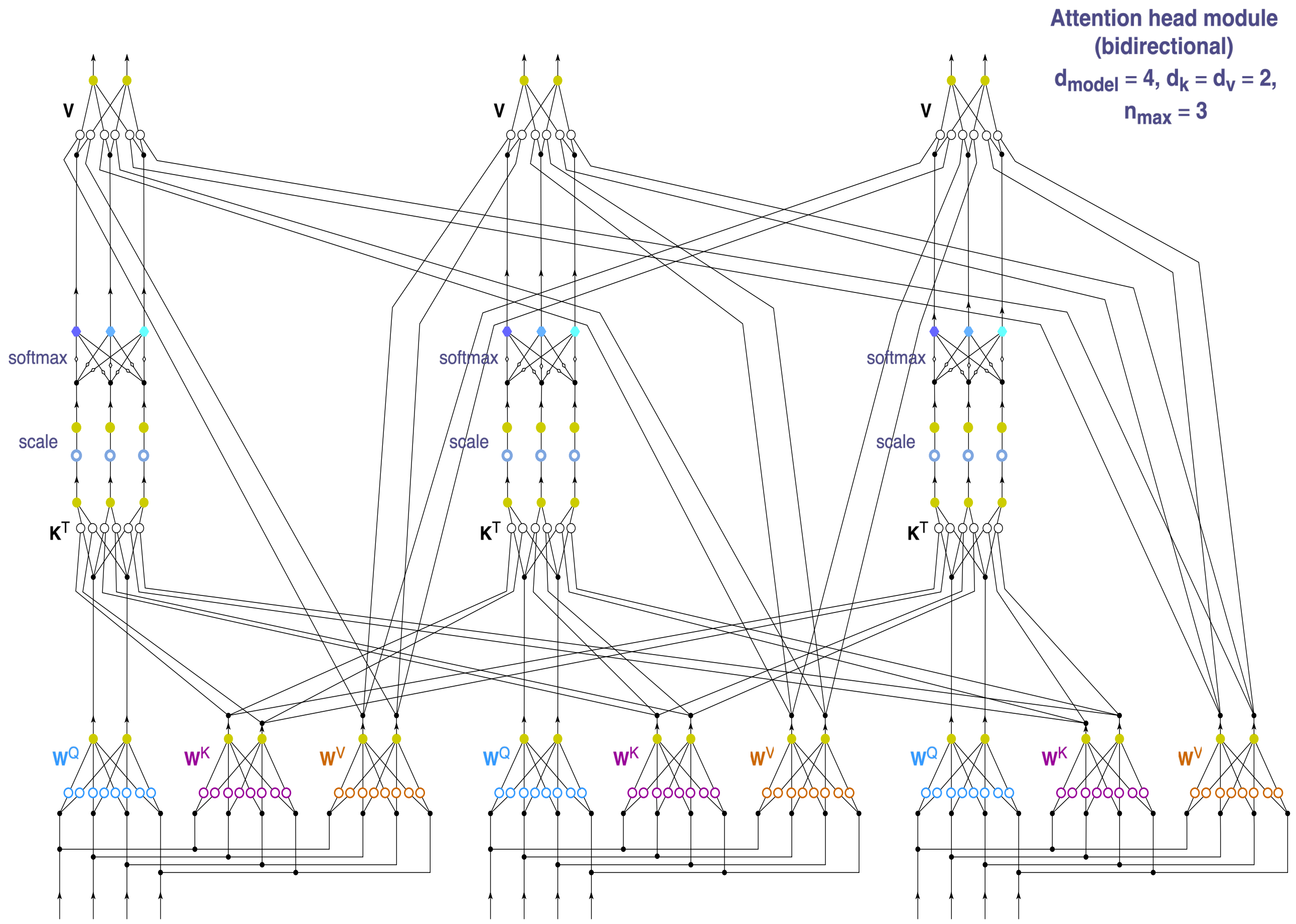

Figure S1. Neuromorphic mechanical realization of a bidirectional attention head.

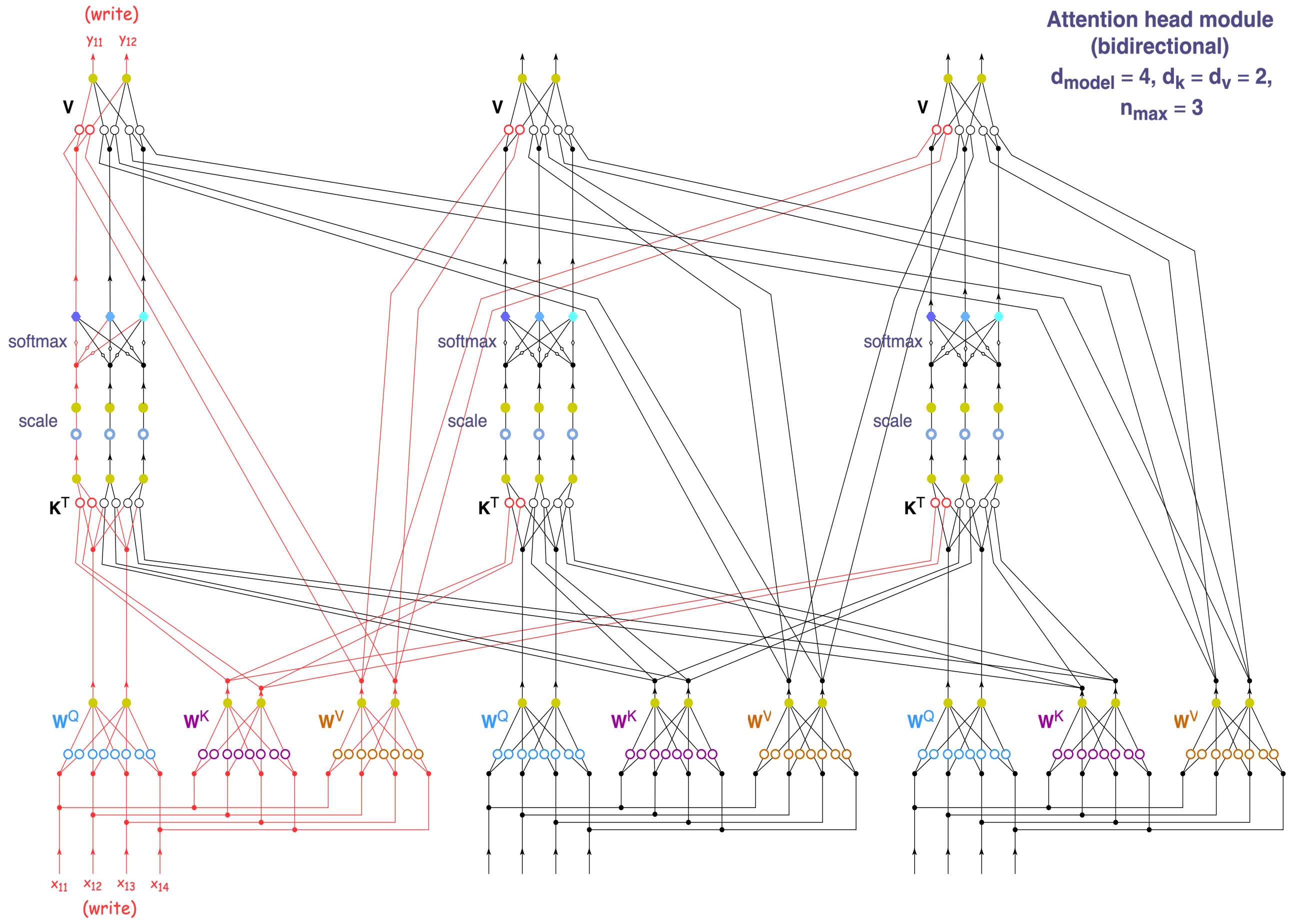

Figure S1a. The same realization activated by an input of $n = 1$ token and the corresponding output.

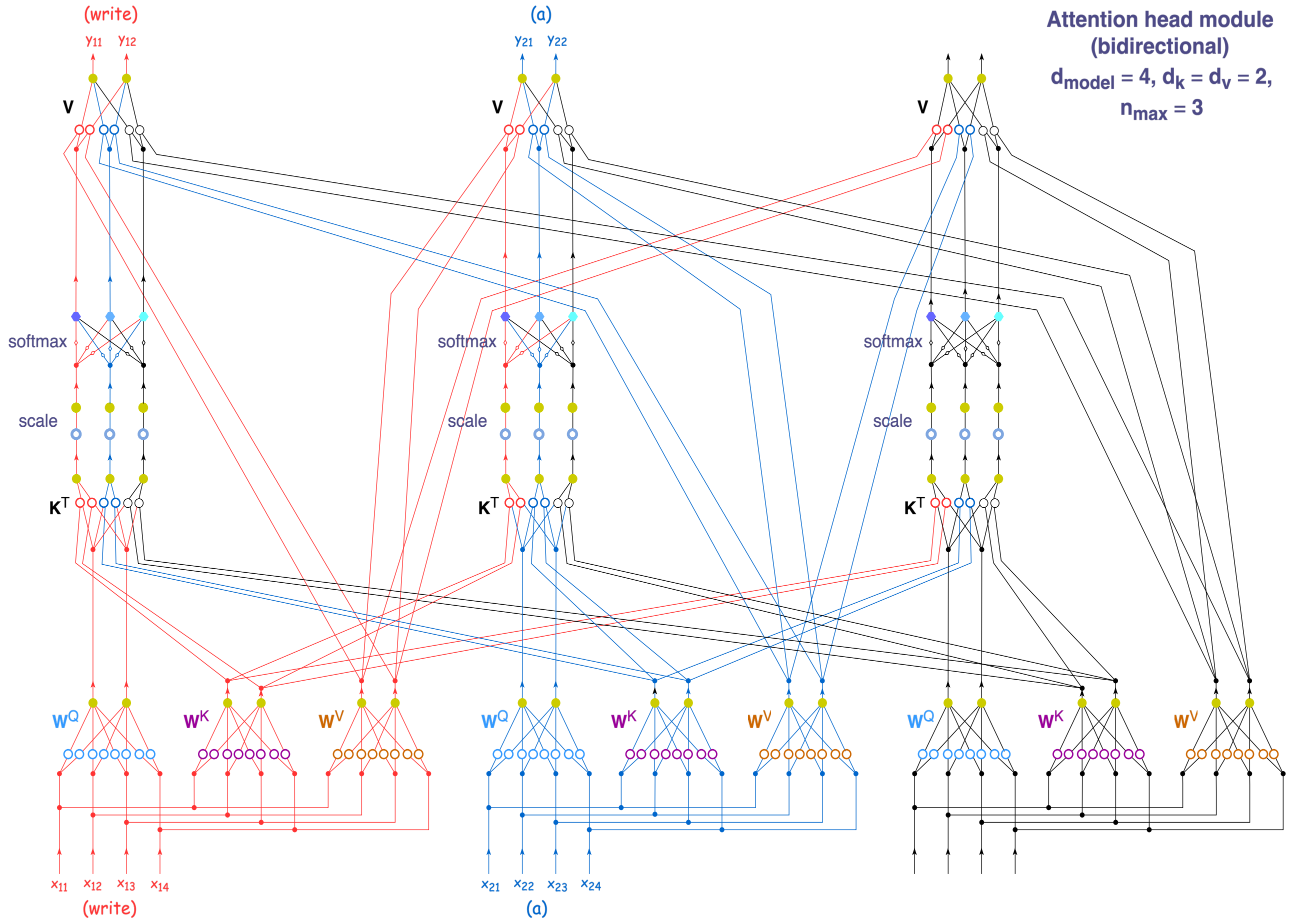

Figure S1b. The same realization activated by an input of $n = 2$ tokens and the corresponding output.

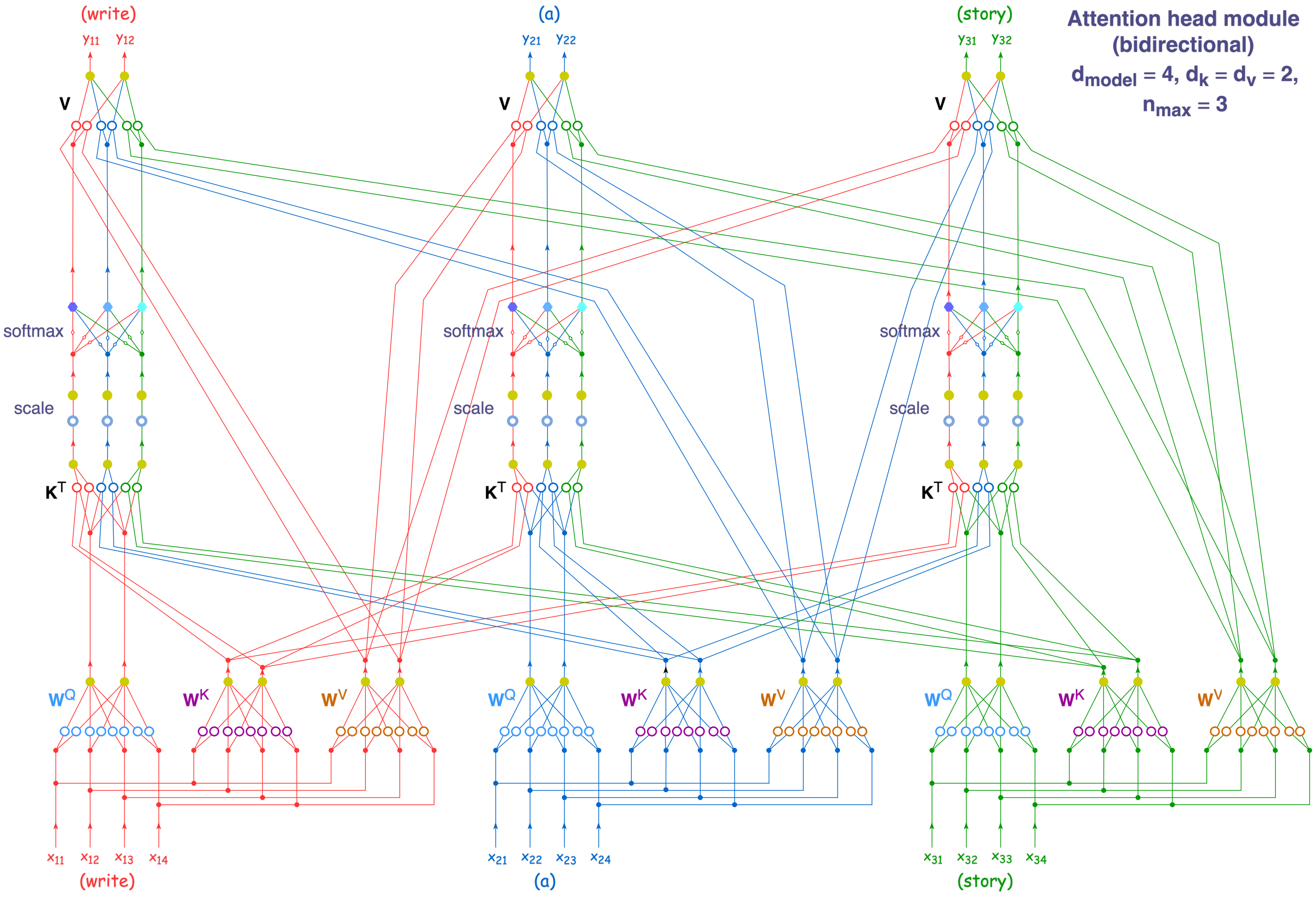

Figure S1c. The same realization activated by an input of $n = 3$ tokens and the corresponding output.

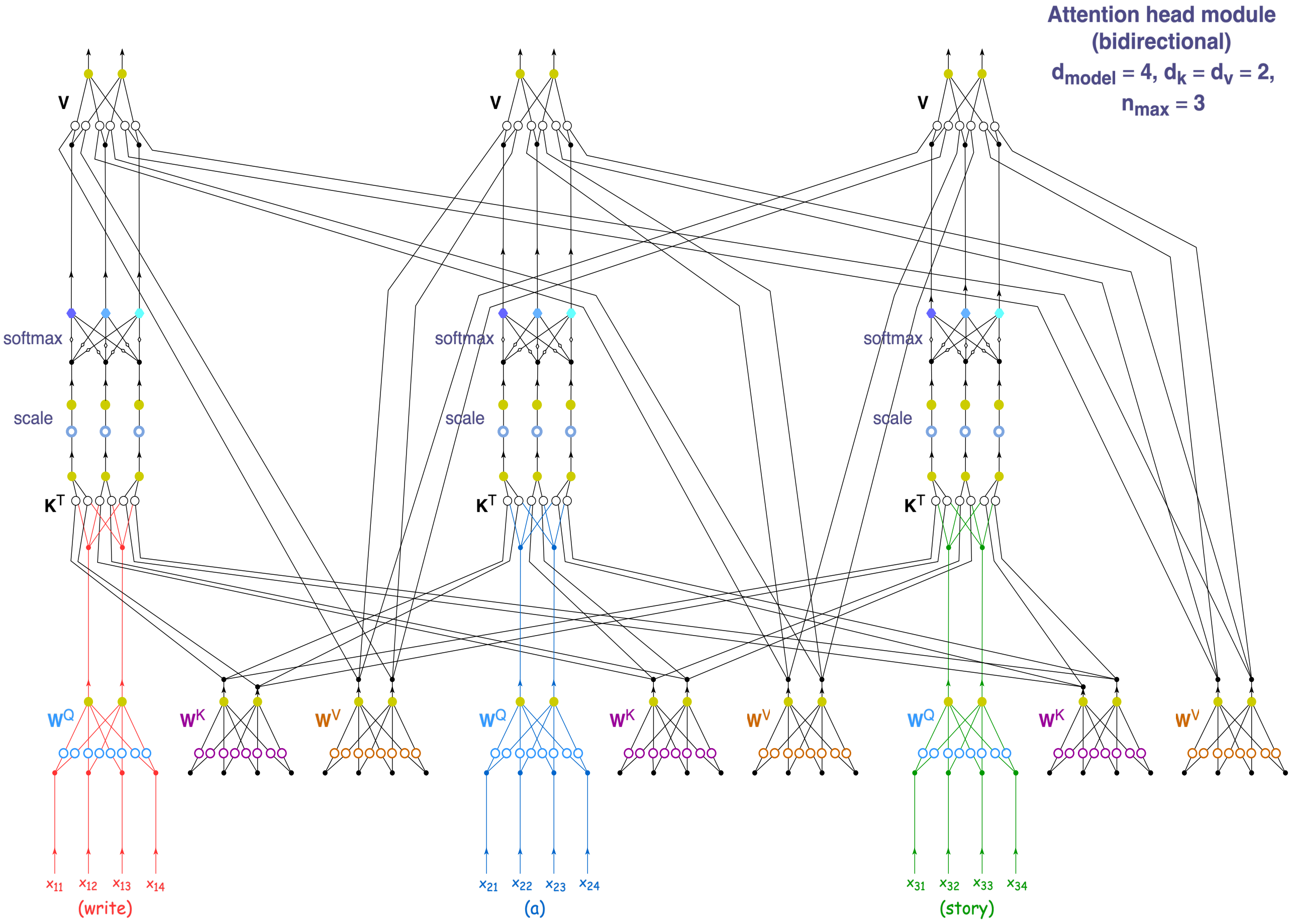

Figure S1d. The same realization with an input of $n = 3$ tokens supplied only to the networks generating queries. No output is produced by the attention head because the $K^\top$ and $V$ networks receive no dynamically generated weights and therefore cannot process any input.

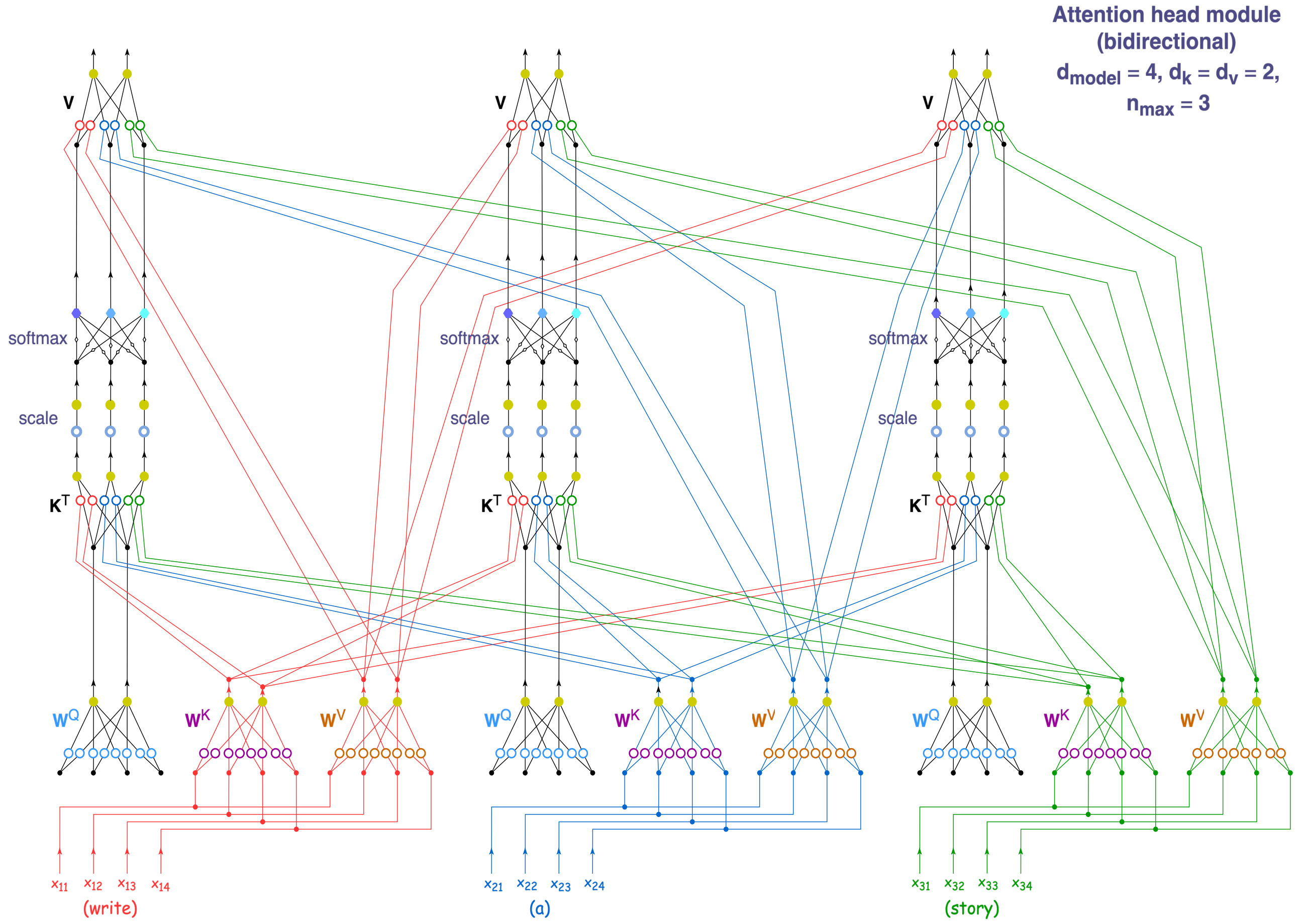

Figure S1e. The same realization with an input of $n = 3$ tokens supplied only to the networks generating keys and values. Their signals generate dynamic weights through output-weight interconnections, and no output is produced by the attention head.

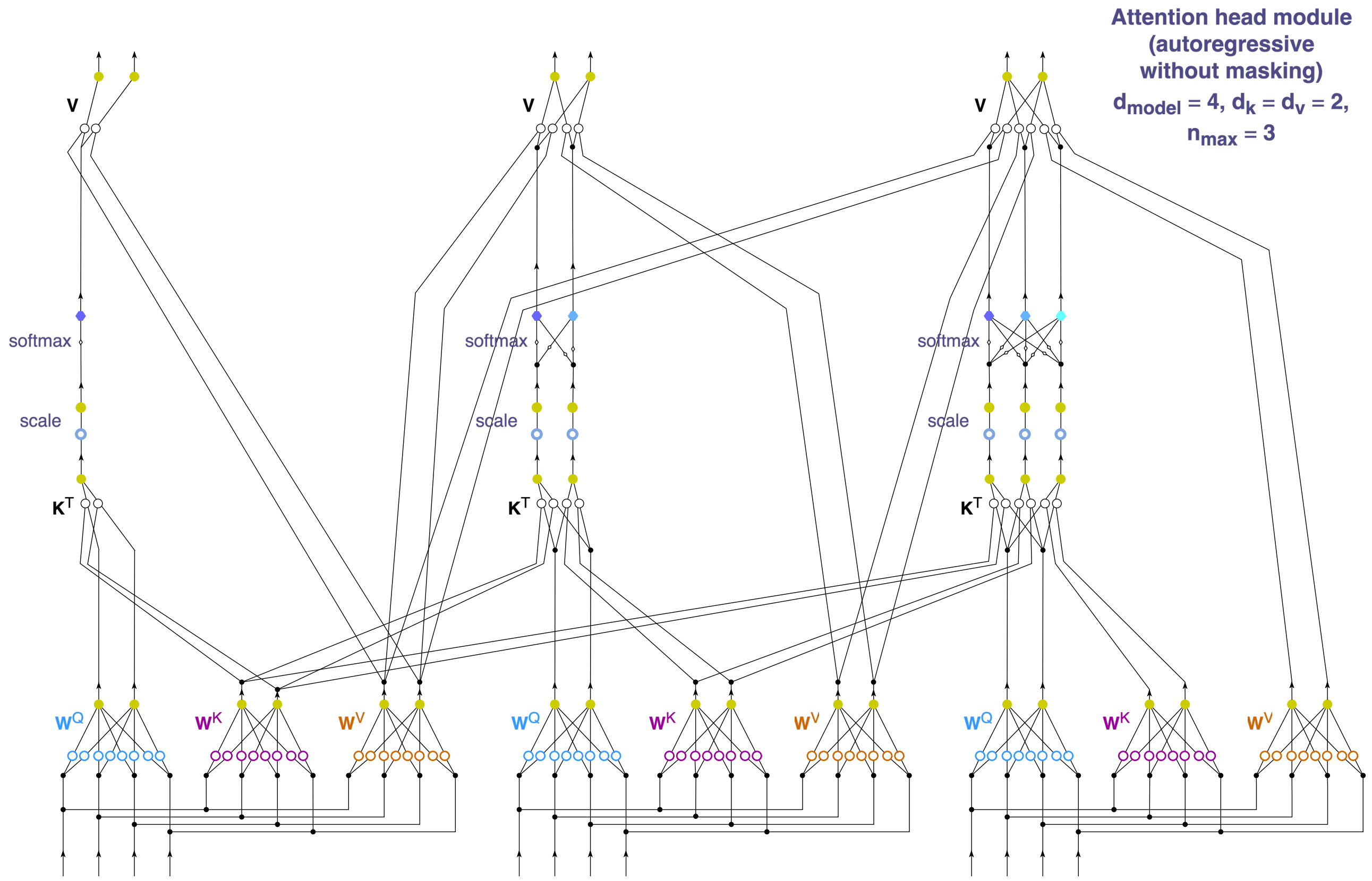

Figure S2. Neuromorphic mechanical realization of an autoregressive attention head without masking.

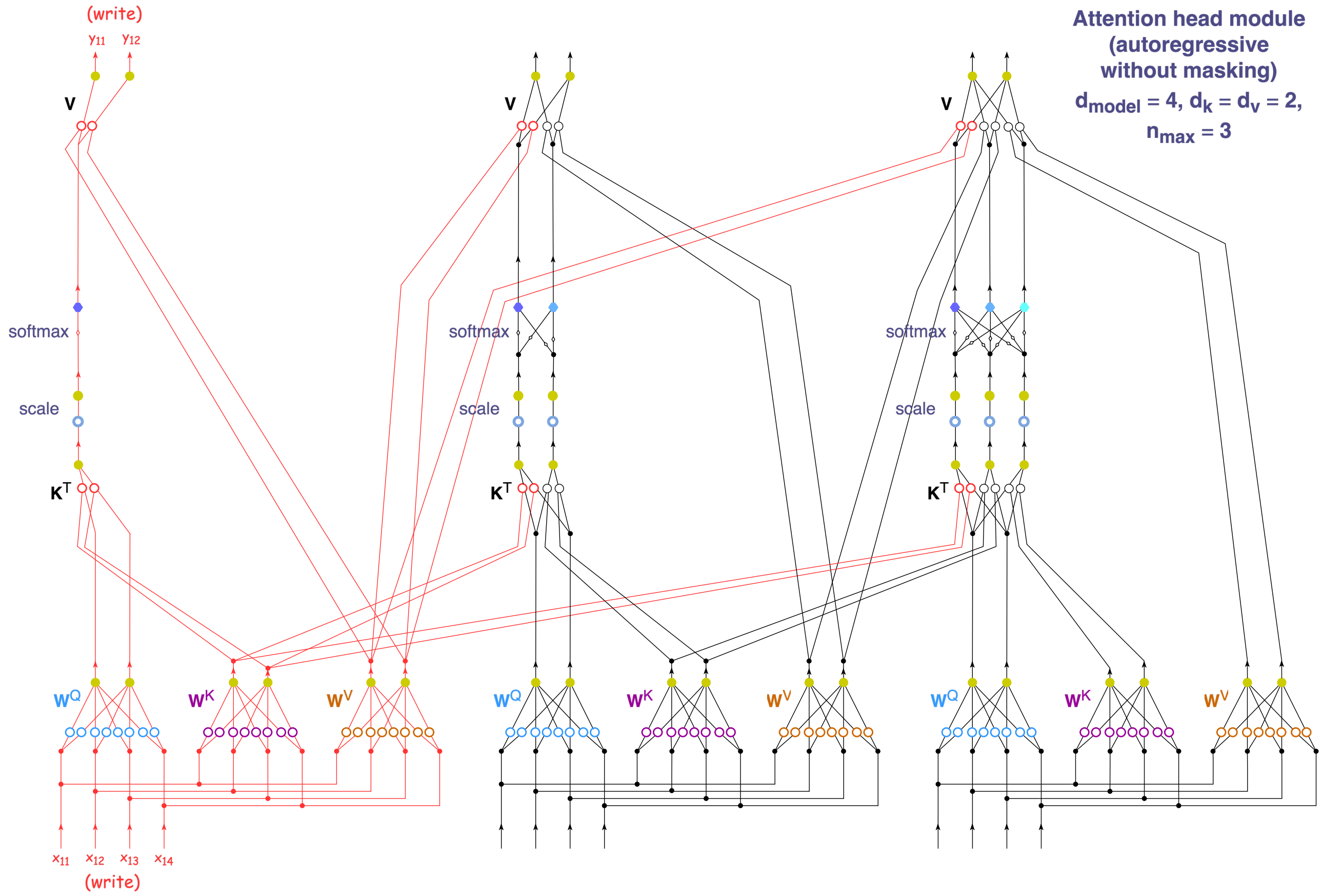

Figure S2a. The same realization activated by an input of $n = 1$ token and the corresponding output.

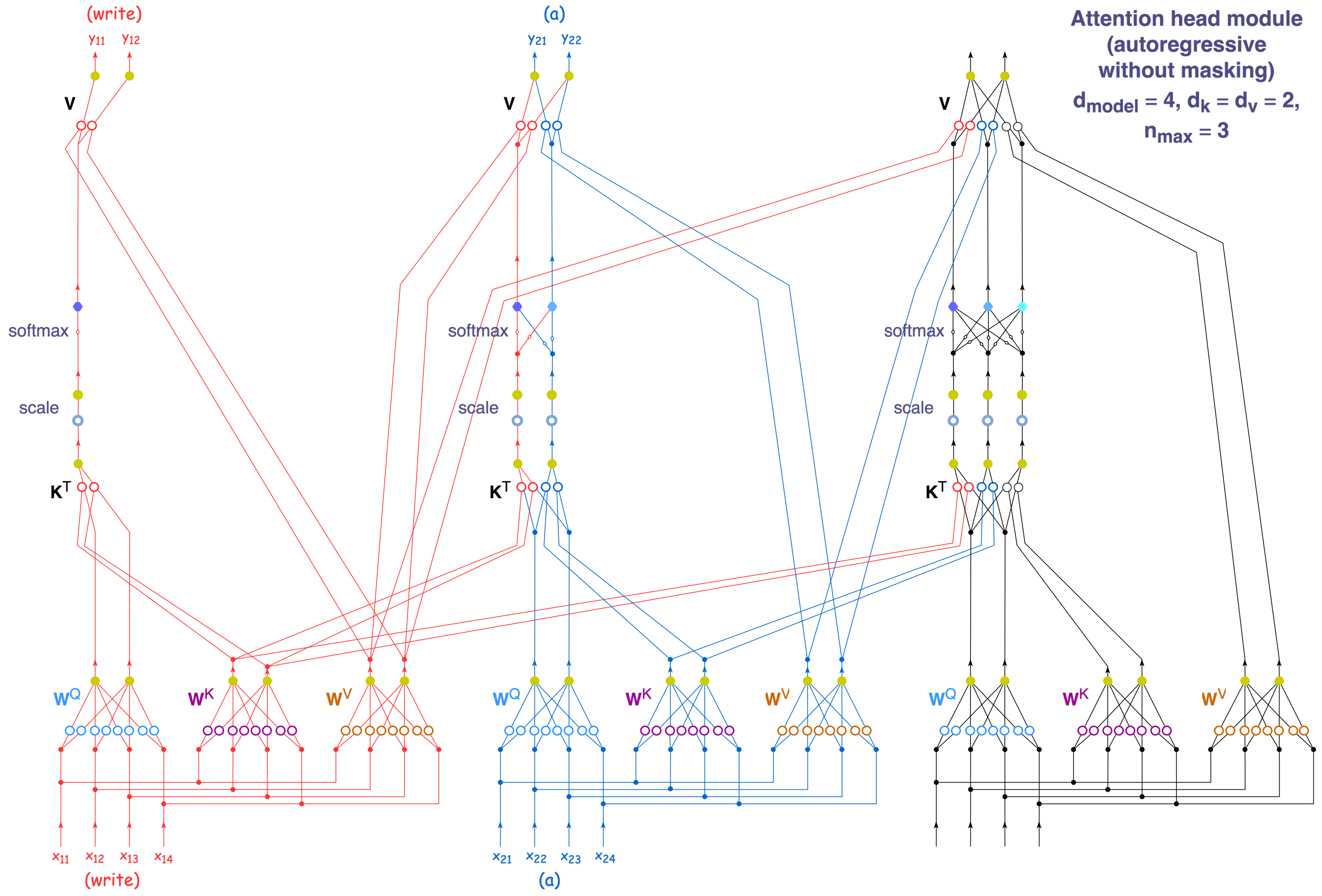

Figure S2b. The same realization activated by an input of $n = 2$ tokens and the corresponding output.

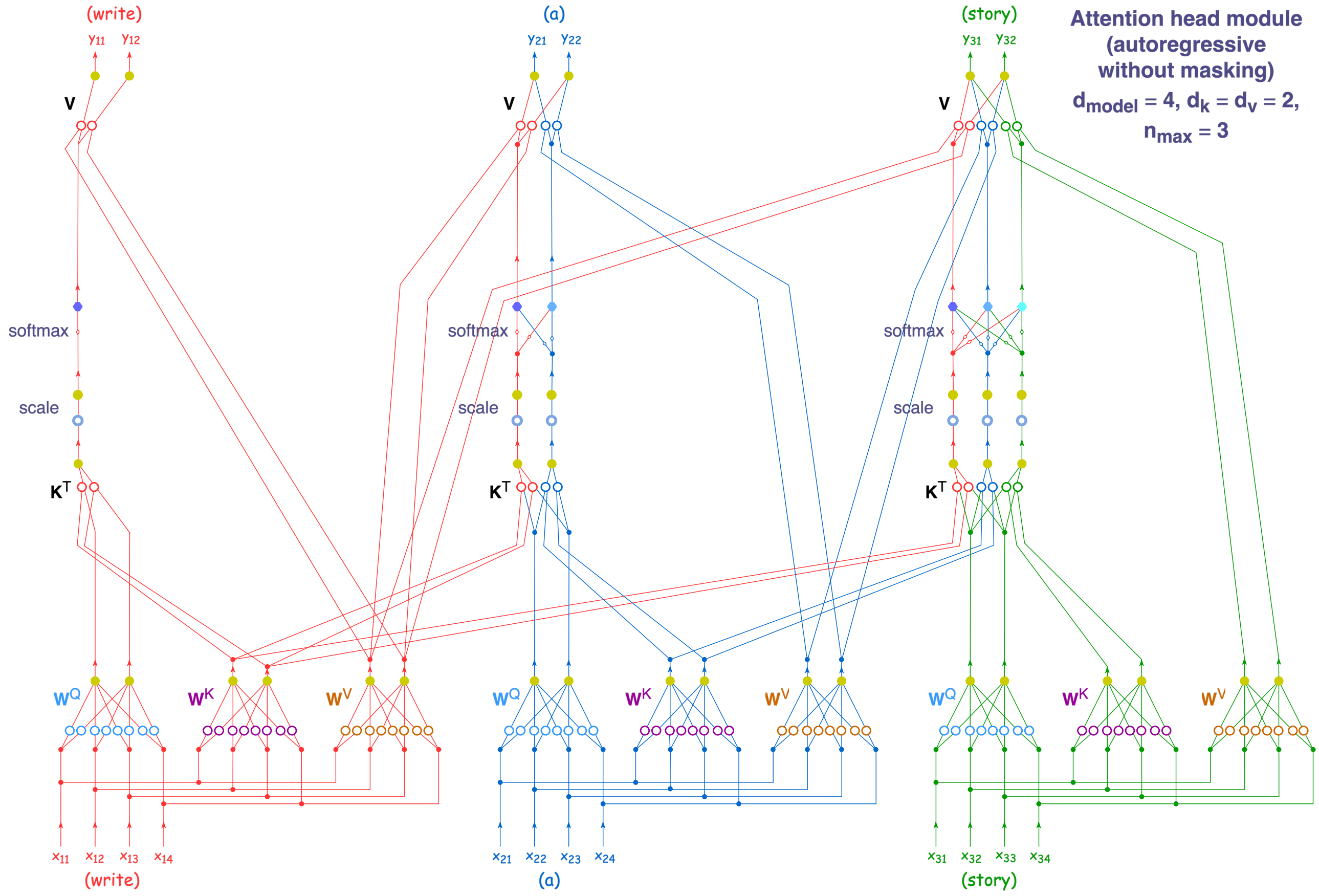

Figure S2c. The same realization activated by an input of $n = 3$ tokens and the corresponding output.

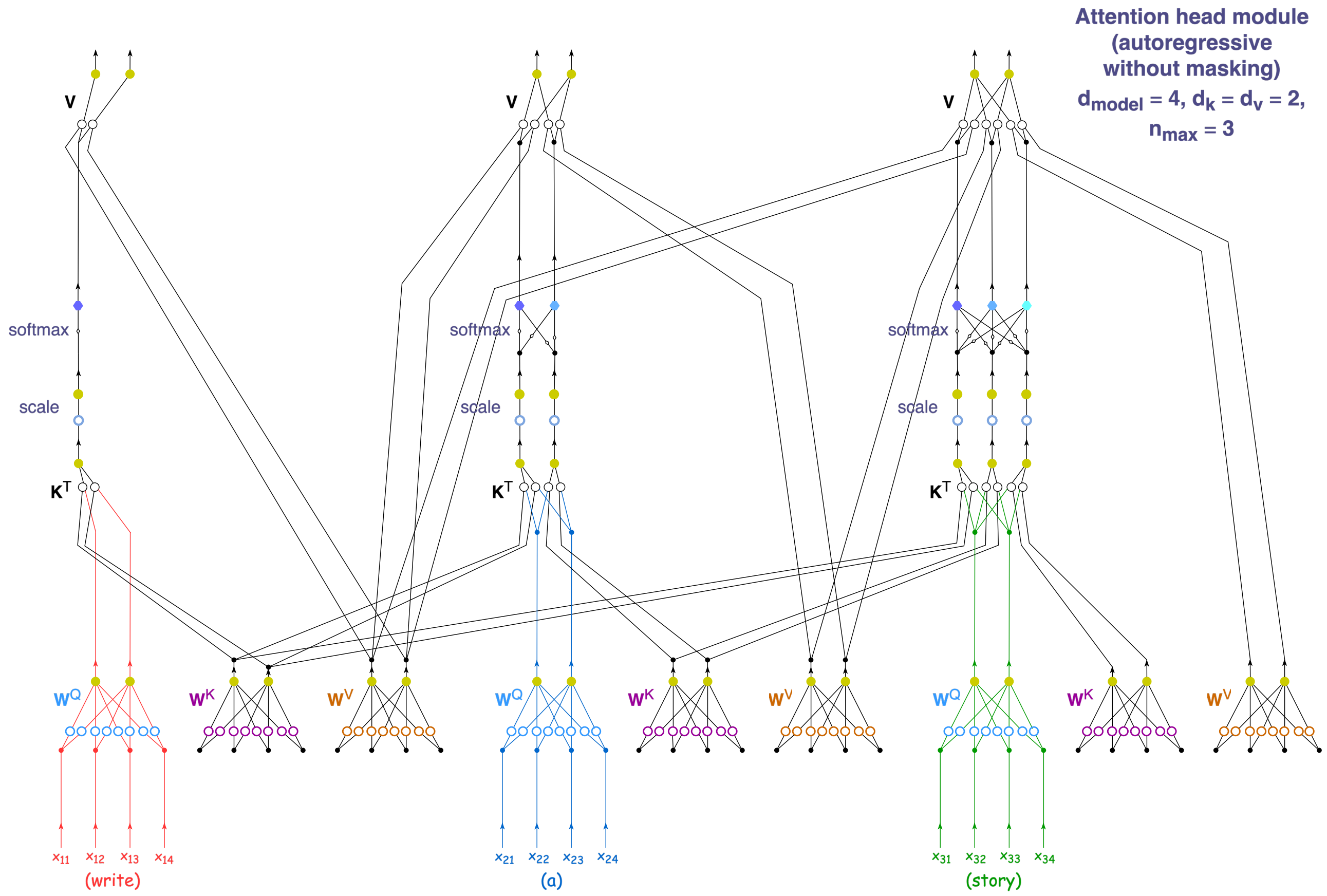

Figure S2d. The same realization with an input of $n = 3$ tokens supplied only to the networks generating queries. No output is produced by the attention head because the $K^{\top}$ and $V$ networks receive no dynamically generated weights and therefore cannot process any input.

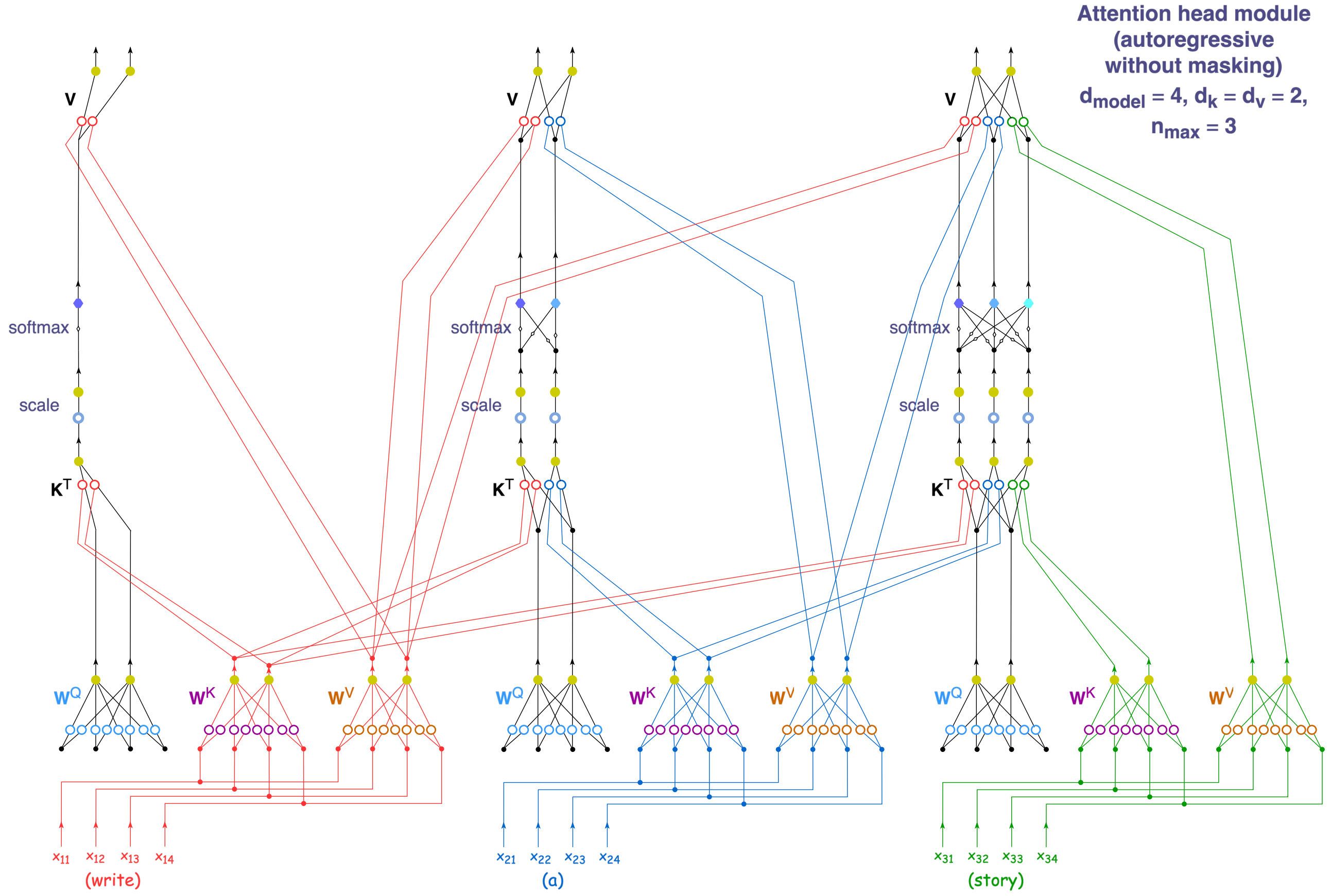

Figure S2e. The same realization with an input of $n = 3$ tokens supplied only to the networks generating keys and values. Their signals generate dynamic weights through output-weight interconnections, and no output is produced by the attention head.